\documentclass[12pt]{article}

\usepackage{amssymb,amsmath,epsfig,cite}

\normalbaselines

\begin{document}

\title{Supersymmetric Quantum Mechanics}

\author{{\bf David J. Fern\'andez C.} \\ Depto. de F\'{\i}sica, Cinvestav \\ A.P. 14-740, 07000 M\'exico D.F., Mexico}

\date{}

\maketitle

\begin{abstract}
Supersymmetric quantum mechanics (SUSY QM) is a powerful tool for generating new potentials with
known spectra departing from an initial solvable one. In these lecture notes we will present some
general formulae concerning SUSY QM of first and second order for one-dimensional arbitrary
systems, and we will illustrate the method through the trigonometric P\"oschl-Teller potentials.
Some intrinsically related subjects, as the algebraic structure inherited by the new Hamiltonians
and the corresponding coherent states will be analyzed. The technique will be as well implemented
for periodic potentials, for which the corresponding spectrum is composed of allowed bands
separated by energy gaps.
\end{abstract}

\maketitle


\section{Introduction}

The techniques based on the factorization method, which try to identify the class of Hamiltonians
whose spectral problem can be algebraically solved, have attracted for years people's attention.
In these lecture notes we will elaborate a brief survey of this subject; since we shall show that
the factorization method, supersymmetric quantum mechanics (SUSY QM), and intertwining technique
are equivalent procedures for generating new solvable potentials departing from an initial one,
these names will be indistinctly used to characterize the technique we are interested in. For a
collection of books and review articles concerning factorization method and related subjects the
reader can seek \cite{ih51}-\cite{do07}.

I would like to emphasize that these lecture notes on SUSY QM are a sequel of those delivered by
myself in 2004 at the Latin-American School of Physics \cite{ff05}. For people wanting to have a
global view of the way our group addresses the subject, and to know more details of our
achievements, I recommend to read both, this article and \cite{ff05}, because they complement to
each other.

From a historical viewpoint, it is nowadays accepted that the factorization method was introduced
for the first time by Dirac in 1935 in order to derive algebraically the spectrum of the harmonic
oscillator \cite{di35}. In coordinates such that $\hbar=\omega=m=1$ the oscillator Hamiltonian
admits the following factorizations:
\begin{eqnarray}
&& \hskip 3.5cm H = a a^+  - \frac12 = a^+  a + \frac12, \\
&& H = - \frac{1}{2}\frac{d^2}{dx^2} + \frac{x^2}{2}, \quad
a = \frac{1}{\sqrt{2}}\left(\frac{d}{dx} + x \right), \quad a^+ =
\frac{1}{\sqrt{2}}\left( -\frac{d}{dx} + x \right).
\end{eqnarray}
From these equations it is straightforward to derive the intertwining relationships,
\begin{eqnarray}
& H a = a (H-1), \quad H a^+  = a^+  (H+1),
\end{eqnarray}
which lead to the commutators ruling the standard Heisenberg-Weyl algebra:
\begin{eqnarray}
& [H,a] = - a, \quad [H,a^+] = a^+, \quad [a,a^+] = 1.
\end{eqnarray}
These expressions and the fact that $H$ is positive definite allow to generate the eigenfunctions
$\psi_n(x)$ and corresponding eigenvalues $E_n = n + 1/2, \ n = 0,1,\dots$ of $H$.

The second important advance in the subject was done by Schr\"odinger in 1940, who realized that
the factorization method can be as well applied to the radial part of the Coulomb problem
\cite{sch40,sch41}. In this case the explicit factorizations in appropriate units read:
\begin{eqnarray}
&& H_\ell = a_\ell^+  a_\ell - \frac1{2\ell^2} =
a_{\ell+1} a_{\ell+1}^+  - \frac1{{2(\ell+1)}^2}, \\
&& \hskip1cm H_\ell = - \frac12 \frac{d^2}{dr^2} + \frac{\ell (\ell +
1)}{2r^2} - \frac{1}{r}, \\
&& \hskip-1cm a_\ell = \frac{1}{\sqrt{2}} \left(\frac{d}{dr} +
\frac{\ell}{r} - \frac{1}{\ell}\right), \quad a_\ell^+ =
\frac{1}{\sqrt{2}} \left( - \frac{d}{dr} + \frac{\ell}{r} -
\frac{1}{\ell}\right) .
\end{eqnarray}
As in the previous case, these factorizations imply some intertwining relationships:
\begin{eqnarray}
&& H_{\ell+1} a_{\ell+1}^+  = a_{\ell+1}^+  H_{\ell}, \quad
H_{\ell-1} a_{\ell} = a_{\ell} H_{\ell},
\end{eqnarray}
from which it can be derived once again the radial eigenfunctions $R_{n\ell}(r)$ and eigenvalues
$E_n = - 1/(2n^2), \ n = \ell + 1,\ell + 2,\dots$ for the Hamiltonian $H_\ell$.

In a set of works starting from 1941, Infeld and collaborators put forward the method
\cite{in41,hi48}, culminating their research with a seminal paper in which it was classified a
wide set of Schr\"odinger-type Hamiltonians (indeed four families) solvable through the
factorization method \cite{ih51}. The main tool for performing this classification were the
factorizations
\begin{eqnarray}
&& H_m = A_m^+  A_m + \epsilon_m = A_{m+1}
A_{m+1}^+  + \epsilon_{m+1}, \\
&& \hskip1cm H_m = - \frac12 \frac{d^2}{dx^2} + V_m(x), \\
&& \hskip-1cm A_m = \frac{1}{\sqrt{2}} \left[\frac{d}{dx} +
\beta_m(x)\right], \quad A_m^+ = \frac{1}{\sqrt{2}} \left[ -
\frac{d}{dx} + \beta_m(x)\right],
\end{eqnarray}
which lead straightforwardly to the intertwining relationships
\begin{eqnarray}
& H_{m + 1} A_{m + 1}^+  = A_{m + 1}^+  H_{m}, \quad H_{m - 1} A_{m}
= A_{m} H_{m}.
\end{eqnarray}
With these ingredients, Infeld and Hull were able to derive elegantly the eigenfunctions and
eigenvalues for the hierarchy of Hamiltonians $H_m$ \cite{ih51}. Moreover, the very technique
allowed them to determine the several forms of the potentials $V_m(x)$ admitting the factorization
treatment. In this way, after \cite{ih51} the idea that the factorization methods was exhausted
started to dominate; indeed, at some time it was thought that for a potential $V(x)$ to be
solvable through factorization, it just has to appear in the Infeld-Hull classification.

Revealing himself against this idea, in 1984 Mielnik performed a simple generalization of the
factorization method \cite{mi84}. He proposed to look for the most general first-order
differential operators $b, b^+$ factorizing the harmonic oscillator Hamiltonian in the way:
\begin{eqnarray}
&& H = b b^+  - \frac12, \quad
b = \frac{1}{\sqrt{2}}\left[\frac{d}{dx} + \beta(x) \right], \quad b^+ =
\frac{1}{\sqrt{2}}\left[ -\frac{d}{dx} + \beta(x) \right].
\end{eqnarray}
Mielnik was able to determine the most general form of $\beta(x)$. Moreover, it turns out that the
product $b^+ b$ supplies a new Hamiltonian
\begin{eqnarray}
&& \widetilde H = b^+  b + \frac12 = - \frac12\frac{d^2}{dx^2} +
\widetilde V(x),
\\ && \hskip1cm \widetilde V(x) = \frac{x^2}2 - \beta'(x) + 1 .
\end{eqnarray}
Hence, the following intertwining relationships are valid:
\begin{eqnarray}
& \widetilde H b^+  = b^+  (H+1), \quad H b = b (\widetilde H - 1),
\end{eqnarray}
which interrelate the eigenfunctions of the harmonic oscillator Hamiltonian $H$ with those of
$\widetilde H$ and vice versa.

Mielnik's work was a breakthrough in the development of the factorization method, since it opened
new ways to explore exactly solvable potentials in quantum mechanics. In particular, his
generalization was quickly applied to the radial part of the Coulomb problem \cite{fe84}. The
important resultant expressions for that system become:
\begin{eqnarray}
&& H_\ell = A_\ell^+  A_\ell - \frac1{2\ell^2}, \quad \widetilde
H_{\ell - 1} = A_{\ell} A_{\ell}^+  - \frac1{2\ell^2}, \\
&& \hskip0.5cm  \widetilde H_{\ell - 1} A_\ell = A_\ell H_\ell, \quad H_\ell
A_\ell^+  = A_\ell^+  \widetilde H_{\ell - 1}.
\end{eqnarray}
They allow to generate the eigenfunctions and eigenvalues of $\widetilde H_{\ell - 1}$ departing
from those of $H_\ell$. Notice that, in the same year the connections between the factorization
method with the Darboux transformation and Witten's supersymmetric quantum mechanics were
established by Andrianov et al \cite{abi84a,abi84b} and Nieto \cite{ni84} respectively.

In 1985 Sukumar \cite{su85a,su85b} pushed further Mielnik's factorization by applying it to
general one-dimensional potentials $V(x)$ and arbitrary factorization energies $\epsilon$ in the
way
\begin{eqnarray}
&& \hskip1.5cm  H = A A^+ + \epsilon, \quad
\widetilde H = A^+ A  + \epsilon, \\
&& \hskip0.5cm H = - \frac12 \frac{d^2}{dx^2} + V(x), \quad \widetilde H = -
\frac12 \frac{d^2}{dx^2} + \widetilde V(x), \\
&& A = \frac{1}{\sqrt{2}} \left[\frac{d}{dx} + \beta(x)\right], \ \
A^+ = \frac{1}{\sqrt{2}} \left[ - \frac{d}{dx} + \beta(x)\right].
\end{eqnarray}
These relationships imply that
\begin{eqnarray}
& \widetilde H A^+ = A^+ H, \quad H A = A \widetilde H , \label{intertwining}
\end{eqnarray}
which allow to generate the eigenfunctions and eigenvalues of $\widetilde H$ from those of $H$.

Along the years, several particular cases and applications of the generalized factorization of
\cite{mi84,fe84,su85a} have been performed successfully \cite{lp86}-\cite{qu08}. Up to this point,
however, the factorization operators, becoming the intertwiners in the equivalent formalism, were
differential operators of first order. A natural generalization, posed in 1993 by Andrianov and
collaborators \cite{ais93,aicd95}, requires that Eq.(\ref{intertwining}) be valid but $A, A^+ $
are substituted by differential operators of order greater than one. Through this higher-order
case, it is possible to surpass the restriction arising from the first-order method that we can
only modify the ground state energy of the initial Hamiltonian. In 1995 Bagrov and Samsonov
proposed the same generalization, and they offered an alternative view to this issue \cite{bs95}.
On the other hand, our group arrived to the subject in 1997 \cite{fe97}-\cite{crf01}, although
there were several previous works related somehow to this generalization
\cite{fe84a}-\cite{fno96}. It is worth to mention explicitly some important contributions that
members of our group have made to the factorization method and related subjects.

\begin{itemize}

\item Generation of SUSY partner of the harmonic oscillator \cite{mi84,fgn98,fhm98}

\item Construction of coherent states for SUSY partners of the harmonic oscillator
\cite{fhn94,fnr95,ro96,fh99} and for general one-dimensional Hamiltonians \cite{fhr07}.

\item Determination of SUSY partners of the radial oscillator potential \cite{fno96}

\item Characterization of systems ruled by polynomial Heisenberg algebras and connection with SUSY
partners of the oscillator \cite{fh99,fnn04,cfnn04,mn08}

\item Analysis of SUSY partners of the radial Coulomb problem \cite{fe84,ro98a,ro98b,fs05}

\item Implementation of SUSY transformations involving complex factorization `energies'
\cite{fmr03}-\cite{fr08}

\item Study of the confluent algorithm for second-order SUSY transformations and applications
\cite{fs05,mnr00,fs03}

\item Generation of SUSY partners of the  P\"oschl-Teller potentials \cite{dnnr99}-\cite{cf08}

\item Application of SUSY techniques to the Lam\'e and associated Lam\'e potentials
\cite{fnn00}-\cite{fg07}

\end{itemize}

It is important to mention that some other groups have elaborated the same subject from different
viewpoints, e.g., $N$-fold supersymmetry by Tanaka and collaborators \cite{ast01}-\cite{bt09},
hidden nonlinear supersymmetries by Plyushchay et al \cite{lp03}-\cite{cjp08}, etc.

\section{Standard supersymmetric quantum mechanics}

In this section we are going to show that the factorization method, intertwining technique and
supersymmetric quantum mechanics are equivalent procedures for generating new solvable potentials
departing from an initial one. The most straightforward way to see this equivalence is starting
from the technique which involves just first-order differential intertwining operators.

\subsection{Intertwining technique}

Suppose we have two Schr\"odinger Hamiltonians
\begin{eqnarray}
&& H_0 = - \frac12\frac{d^2}{dx^2} + V_0(x), \quad
H_1 = - \frac12\frac{d^2}{dx^2} + V_1(x),
\end{eqnarray}
which are intertwined by first-order differential operators $A_1,A_1^+$ in the way
\begin{eqnarray}
&& \hskip2cm H_1 A_1^+ = A_1^+ H_0, \qquad H_0 A_1 = A_1 H_1, \label{inter1gen} \\
&& A_1 = \frac{1}{\sqrt{2}}\left[\frac{d}{dx}
+ \alpha_1(x)\right], \quad  A_1^+ = \frac{1}{\sqrt{2}}\left[-\frac{d}{dx} +
\alpha_1(x)\right].
\end{eqnarray}
By employing that, at the operator level, it is valid
\begin{eqnarray}
&& \frac{d}{dx}f(x) = f(x)\frac{d}{dx} + f'(x), \quad \frac{d^2}{dx^2}f(x) =
f(x)\frac{d^2}{dx^2} + 2 f'(x)\frac{d}{dx} + f''(x),
\end{eqnarray}
where $f(x)$ is a multiplicative operator when acting on a generic wavefunction $\psi(x)$, the
calculation of each member of the first intertwining relationship in Eq.(\ref{inter1gen}) leads to
\begin{eqnarray}
&& \sqrt{2}H_1 A_1^+ =  \frac12\frac{d^3}{dx^3} -\frac{\alpha_1}{2}
\frac{d^2}{dx^2} - (V_1+\alpha_1') \frac{d}{dx}
 + \alpha_1 V_1 - \frac{\alpha_1''}{2}, \\
&& \hskip0.8cm \sqrt{2}A_1^+ H_0 =  \frac12\frac{d^3}{dx^3} - \frac{\alpha_1}{2}
\frac{d^2}{dx^2} - V_0 \frac{d}{dx}+ \alpha_1 V_0 - V_0' .
\end{eqnarray}
Thus:
\begin{eqnarray}
&& \hskip1cm V_1 = V_0 - \alpha_1', \label{gv1} \\
&& \alpha_1 V_1 - \frac{\alpha_1''}{2} = \alpha_1 V_0 - V_0' \label{derricc} .
\end{eqnarray}
By plugging the $V_1$ of Eq.(\ref{gv1}) in Eq.(\ref{derricc}) and integrating the result we get:
\begin{eqnarray}
& \alpha_1' + \alpha_1^2 = 2[V_0(x)- \epsilon] .
\end{eqnarray}
This nonlinear first-order differential equation for $\alpha_1$ is the well-known {\it Riccati
equation}.  In terms of a {\it seed Schr\"odinger solution} $u^{(0)}$ such that $\alpha_1 =
{u^{(0)}}'/u^{(0)}$ it turns out that:
\begin{eqnarray}
&&  - \frac12 {u^{(0)}}'' + V_0 u^{(0)} = H_0 u^{(0)} = \epsilon u^{(0)} , \label{seg}
\end{eqnarray}
which is the initial stationary Schr\"odinger equation. These equations allow to establish the
link between the intertwining technique and the {\it factorization method}, since it is valid:
\begin{eqnarray}
&& H_0 = A_1 A_1^+ + \epsilon, \quad H_1 = A_1^+ A_1 + \epsilon. \label{genfac}
\end{eqnarray}

Suppose now that $H_0$ is a given solvable Hamiltonian, with known eigenfunctions $\psi_n^{(0)}$
and eigenvalues $E_n$, namely,
\begin{eqnarray}
& H_0 \psi_n^{(0)} = E_n \psi_n^{(0)}, \quad n=0,1,\dots
\end{eqnarray}
In addition, let us take a nodeless mathematical eigenfunction $u^{(0)}$ associated to
$\epsilon\leq E_0$ in order to implement the intertwining procedure. The equations
(\ref{inter1gen}) and the resulting factorizations (\ref{genfac}) indicate that, if $A_1^+
\psi_n^{(0)} \neq 0$, then $\big\{\psi_n^{(1)} = A_1^+\psi_n^{(0)}/\sqrt{E_n - \epsilon} \big\}$
is an orthonormal set of eigenfunctions of $H_1$ with eigenvalues $\{E_n\}$. This is a basis if
there is not a square-integrable function $\psi_{\epsilon}^{(1)}$ which is orthogonal to the full
set. Thus, let us look for $\psi_{\epsilon}^{(1)}$ such that
\begin{eqnarray}
& 0 = (\psi_{\epsilon}^{(1)},\psi_n^{(1)}) \propto
(\psi_{\epsilon}^{(1)},A_1^+ \psi_n^{(0)}) = (A_1\psi_{\epsilon}^{(1)},\psi_n^{(0)}) \quad \Rightarrow
\quad A_1 \psi_{\epsilon}^{(1)} = 0  .
\end{eqnarray}
By solving this first-order differential equation we get:
\begin{eqnarray}
&& \psi_{\epsilon}^{(1)} \propto e^{-\int_0^x\alpha_1(y) dy}=
\frac{1}{u^{(0)}} .
\end{eqnarray}
Since $H_1 \psi_{\epsilon}^{(1)} = \epsilon \psi_{\epsilon}^{(1)}$ (compare Eq.(\ref{genfac})),
then the spectrum of $H_1$ depends of the normalizability of $\psi_{\epsilon}^{(1)}$. We can
identify three different situations.

\smallskip

\noindent (i) Let us choose $\epsilon = E_0$ and $u^{(0)} = \psi_0^{(0)}$, which is nodeless in
the domain of $V_0$, $\alpha_1 = {\psi_0^{(0)}}'/\psi_0^{(0)}$. Thus, $V_1 = V_0 - \alpha_1'$ is
completely determined, and the corresponding Hamiltonian $H_1$ has eigenfunctions and eigenvalues
given by
\begin{eqnarray}
&& \psi_n^{(1)} = \frac{A_1^+\psi_n^{(0)}}{\sqrt{E_n-E_0}}, \quad {\rm Sp}(H_1)=\{E_n, \ n = 1,2,\dots\}.
\end{eqnarray}
Notice that $E_0\not\in$ Sp$(H_1)$ since $\psi_{\epsilon}^{(1)} \propto 1/u^{(0)}$ is not
square-integrable.

\noindent (ii) Let now a nodeless seed solution $u^{(0)}$ for $\epsilon < E_0$ be chosen so that
$\alpha_1 = {u^{(0)}}'/u^{(0)}$ is non-singular. Note that $u^{(0)}$ diverges at both ends of the
$x$-domain $\Rightarrow \psi_{\epsilon}^{(1)} \propto 1/u^{(0)}$ tends to zero at those ends.
Hence, the full set of eigenfunctions and eigenvalues is
\begin{eqnarray}
&& \psi_{\epsilon}^{(1)} \propto \frac{1}{u^{(0)}}, \quad
\psi_n^{(1)} = \frac{A_1^+\psi_n^{(0)}}{\sqrt{E_n-\epsilon}}, \quad {\rm Sp}(H_1)=\{
\epsilon, E_n, \ n = 0,1,2,\dots\}.
\end{eqnarray}

\smallskip

\noindent (iii) Let us employ a solution $u^{(0)}$ with a node at one of the ends of the
$x$-domain for $\epsilon < E_0$. Thus, the transformation induced by $\alpha_1 =
{u^{(0)}}'/u^{(0)}$ is non-singular, $\psi_{\epsilon}^{(1)} \propto 1/u^{(0)}$ diverges at the end
where $u^{(0)}$ tends to zero, and then the eigenfunctions and eigenvalues of $H_1$ become
\begin{eqnarray}
&& \psi_n^{(1)} =
\frac{A_1^+\psi_n^{(0)}}{\sqrt{E_n-\epsilon}}, \quad {\rm
Sp}(H_1)=\{E_n, \ n = 0,1,2,\dots\}.
\end{eqnarray}
This implies that $H_1$ and $H_0$ are isospectral Hamiltonians.

\subsection{Supersymmetric quantum mechanics}

Let us build up now the so-called supersymmetric quantum mechanics, by realizing the Witten
supersymmetry algebra with two generators \cite{wi81}
\begin{eqnarray}
&& [Q_i, H_{\rm ss}]=0, \quad \{ Q_i,Q_j \} = \delta_{ij} H_{\rm
ss}, \quad i,j=1,2 , \label{susyalg}
\end{eqnarray}
in the way
\begin{eqnarray}
Q_1 & = & \!\!\! \frac{Q^+ + Q}{\sqrt{2}}, \quad Q_2 = \frac{Q^+ - Q}{i\sqrt{2}}, \quad
Q =  \left(\begin{matrix} 0 & 0 \\ A_1 & 0 \end{matrix} \right), \quad Q^+ =
\left(\begin{matrix} 0 & A_1^+ \\ 0 & 0 \end{matrix}
\right), \\
H_{\rm ss} &  = & \!\!\! \{ Q, Q^+ \} = \left(\begin{matrix} H^+ & 0 \\ 0 & H^- \end{matrix}
\right) = \left(\begin{matrix} A_1^+ A_1 & 0 \\ 0 &
A_1 A_1^+ \end{matrix} \right) = \left(\begin{matrix} H_1 - \epsilon & 0 \\ 0 & H_0 - \epsilon \end{matrix}
\right) .
\end{eqnarray}
Notice that there is a linear relationship between $H_{\rm ss}$ and $H^p = {\rm Diag}\{H_1,H_0\}$:
\begin{eqnarray}
H_{\rm ss} = (H^p-\epsilon) .
\end{eqnarray}
There are two-fold degenerated levels of $H_{\rm ss}$ [$E_n - E_0$ with $n=1,2,\dots$ for the case
(i) and $E_n - \epsilon$ with $n=0,1,2,\dots$ for cases (ii) and (iii)] which have associated the
two orthonormal eigenvectors
\begin{eqnarray}
& \left(\begin{matrix} 0 \\ \psi_n^{(0)} \end{matrix} \right), \qquad \left(\begin{matrix}
\psi_n^{(1)} \\ 0 \end{matrix} \right) .
\end{eqnarray}
The ground state energy of $H_{\rm ss}$ in cases (i) and (ii) is non-degenerated (and equal to
$0$), with the corresponding eigenstate being given respectively by
\begin{eqnarray}
& \left(\begin{matrix} 0 \\ \psi_0^{(0)} \end{matrix} \right), \qquad \left(\begin{matrix}
\psi_{\epsilon}^{(1)} \\ 0 \end{matrix} \right) .
\end{eqnarray}
In this case it is said that the supersymmetry is unbroken. On the other hand, in case (iii) the
ground state energy ($E_0 - \epsilon$) is doubly degenerated, and then the supersymmetry is
spontaneously broken.

\section{Example: trigonometric P\"oschl-Teller potentials}

Let us apply the previous technique to the trigonometric P\"oschl-Teller potentials
\cite{cf07,cf08}:
\begin{eqnarray}
&& V_0(x) = {(\lambda - 1)\lambda\over 2\sin^2(x)} + {(\nu-1)\nu\over
2\cos^2(x)}, \quad 0\leq x\leq \pi/2, \quad  \lambda,\nu > 1 . \label{ptp}
\end{eqnarray}
First of all, we require the general solution of the Schr\"odinger equation (\ref{seg}) with the
$V_0(x)$ given in (\ref{ptp}), which is given by
\begin{eqnarray}
u^{(0)}(x) & = & \sin^\lambda(x) \cos^\nu(x) \bigg\{ A \
{}_2F_1\left[\frac{\mu}{2} + \sqrt{\frac{\epsilon}{2}},\frac{\mu}{2}
- \sqrt{\frac{\epsilon}{2}}; \lambda + \frac12;\sin^2(x)\right]
\nonumber \\
&& \hskip-1.3cm  + B  \sin^{1-2\lambda}(x)  \, {}_2F_1\left[\frac{1 + \nu
- \lambda}{2} + \sqrt{\frac{\epsilon}{2}},\frac{1 + \nu -
\lambda}{2} - \sqrt{\frac{\epsilon}{2}}; \frac32 -
\lambda;\sin^2(x)\right]\bigg\}, \label{vptgs}
\end{eqnarray}
where $\mu = \lambda + \nu$. The physical eigenfunctions $\psi_{n}^{(0)}(x)$ satisfy
$\psi_{n}^{(0)}(0) = \psi_{n}^{(0)}(\pi/2) = 0$. If we ask that $\psi_{n}^{(0)}(0) = 0$, it turns
out that $B = 0$. Moreover, in order to avoid that the divergent behavior at $x = \pi/2$ of the
hypergeometric function of the remaining term dominates over the null one induced by $\cos^\nu(x)$
for $E>0$ it has to happen that
\begin{eqnarray}
&& \frac{\mu}{2} \pm \sqrt{\frac{E_n}{2}} = - n \quad \Rightarrow \quad
E_{n} =  \frac{(\mu + 2n)^2}2, \quad n=0,1,2,\ldots
\label{eigenenergies}
\end{eqnarray}
The corresponding normalized eigenfunctions are given by
\begin{eqnarray}
&\hskip-0.5cm   \psi_n^{(0)} =  \left[ \frac{2(\mu + 2n)n!\Gamma(\mu + n)(\lambda + \frac12)_n}{
\Gamma(\lambda + \frac12)\Gamma^3(\nu + \frac12)(\nu + \frac12)_n}\right]^{\frac12} \sin^\lambda(x)
\cos^\nu(x) \ {}_2F_1[-n,n  +  \mu; \lambda  + \frac12;\sin^2(x)].
\end{eqnarray}

It will be important later to know the number of zeros which $u^{(0)}$ has, according to the value
taken by $\epsilon$. This information can be obtained by comparing the asymptotic behavior of
$u^{(0)}$ at $x\rightarrow 0$ and $x\rightarrow \pi/2$; by picking up $B = 1$ and $A = -b/a + q$
in Eq.(\ref{vptgs}) with
\begin{eqnarray}
&& a = \frac{\Gamma(\lambda + \frac12) \Gamma(\nu -
\frac12)}{\Gamma(\frac{\mu}{2} + \sqrt{\frac{\epsilon}{2}})
\Gamma(\frac{\mu}{2} - \sqrt{\frac{\epsilon}{2}})}, \quad b =
\frac{\Gamma(\frac32 - \lambda) \Gamma(\nu -
\frac12)}{\Gamma(\frac{1 + \nu - \lambda}{2} +
\sqrt{\frac{\epsilon}{2}}) \Gamma(\frac{1 + \nu - \lambda}{2} -
\sqrt{\frac{\epsilon}{2}})},
\end{eqnarray}
it turns out that:

\begin{itemize}

\item If $\epsilon < E_0$, $u^{(0)}$ has either $0$ or $1$ nodes for $q > 0$ or $q < 0$
respectively

\item In general, if $E_{i-1} <\epsilon < E_i$, $u^{(0)}$ will have either $i$ or $i+1$ nodes for
$q > 0$ or $q < 0$ respectively, $i=2,3,\dots$

\end{itemize}

Concerning the spectral modifications which can be induced through the standard first-order
supersymmetric quantum mechanics, it is obtained the following.

\medskip

\noindent(a) Deleting the ground state of $H_0$. Let us choose $\epsilon = E_0$ and
\begin{eqnarray}
& u^{(0)} = \psi_0^{(0)} \propto \sin^\lambda(x) \cos^\nu(x).
\end{eqnarray}
The first-order SUSY partner of $V_0$ becomes (see Eq.(\ref{gv1}))
\begin{eqnarray}
&& V_1 = {\lambda(\lambda + 1)\over 2\sin^2(x)} + {\nu(\nu +
1)\over 2\cos^2(x)}, \quad \lambda,\nu > 1 . \label{v1ptd}
\end{eqnarray}
Since $\psi_{\epsilon}^{(1)} \propto 1/\psi_0^{(0)}$ diverges at $x = 0,\pi/2$ $\Rightarrow
E_0\not\in {\rm Sp}(H_1)= \{E_n,n=1,2,\dots\}$. Note that $V_1$ can be obtained from $V_0$ by
changing $\lambda\rightarrow \lambda + 1, \ \nu \rightarrow \nu + 1$; this property is known
nowadays as {\it shape invariance} \cite{ge83}, although it was known since Infeld-Hull work
\cite{ih51}.

\smallskip

\noindent(b) Creating a new ground state. Let us take now $\epsilon < E_0$ and a nodeless
$u^{(0)}$ (Eq.(\ref{vptgs}) with $B=1$, $A = -b/a + q$, $q>0$). As $u^{(0)}\rightarrow \infty$
when $x\rightarrow 0, \pi/2$ $\Rightarrow$ $\psi_\epsilon^{(1)} (0) = \psi_\epsilon^{(1)} (\pi/2)
= 0$, i.e., $\psi_\epsilon^{(1)}$ is an eigenfunction of $H_1$ with eigenvalue $\epsilon$, and
therefore Sp($H_1$)=$\{\epsilon, E_n, n=0,1,\dots\}$. In order to deal with the singularities at
$x = 0, \pi/2$ induced by $u^{(0)}$ on $V_1$, it is expressed $u^{(0)} = \sin^{1-\lambda}(x)
\cos^{1-\nu}(x) {\rm v},$ where ${\rm v}$ is a nodeless bounded function in $[0,\pi/2]$. Hence
\begin{eqnarray}
&& V_1  =  {(\lambda - 2)(\lambda - 1)\over
2\sin^2(x)} + {(\nu - 2)(\nu - 1)\over 2\cos^2(x)} - [\ln {\rm v}]'', \quad
\lambda, \nu > 2 . \label{v1ptc}
\end{eqnarray}

\smallskip

\noindent(c) Isospectral potentials. They arise from the previous case (for $\epsilon<E_0$) in the
limit when $u^{(0)}$ acquires a node at one of the ends of the domain so that
$\psi_\epsilon^{(1)}$ leaves to be an eigenstate of $H_1$. We can take, for example, the solution
given in Eq.(\ref{vptgs}) with $A=1$, $B=0$ $\Rightarrow$ $u^{(0)}(0) = 0$. We isolate the
singularities induced on $V_1$ by expressing $u^{(0)} = \sin^{\lambda}(x)\cos^{1-\nu}(x) {\rm v}$,
where ${\rm v}$ is a nodeless bounded function in $[0,\pi/2]$. Hence
\begin{eqnarray}
&& V_1 = {\lambda(\lambda + 1)\over
2\sin^2(x)} + {(\nu - 2)(\nu - 1)\over 2\cos^2(x)} - [\ln {\rm v}]'', \quad
\lambda>1, \nu > 2 . \label{v1pti}
\end{eqnarray}
Notice that now $H_1$ and $H_0$ are isospectral.

An illustration of the several first-order SUSY partner potentials $V_1(x)$ of $V_0(x)$ (see
Eqs.(\ref{v1ptd}-\ref{v1pti})) is given in Figure 1.

\begin{figure}[ht]
\centering \epsfig{file=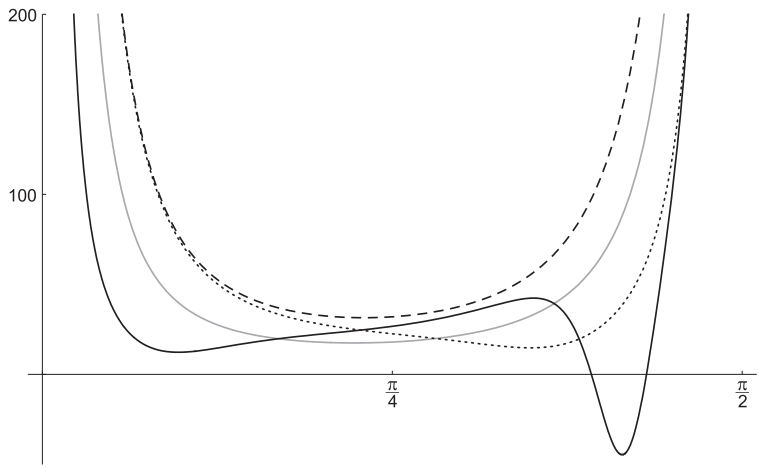, width=10cm}
\caption{Trigonometric P\"oschl-Teller potential $V_0(x)$ for $\lambda = 3, \nu = 4$ (gray curve)
and its first-order SUSY partners $V_1(x)$ generated through the following spectral modifications:
(a) deleting the ground state $E_0 = 24.5$ (dashed curve); (b) creating a new ground state at
$\epsilon = 19$ (continuous curve); (c) isospectral transformation for $\epsilon = 19$ (dotted
curve) \cite{cf08}}
\end{figure}

\section{Second-order Supersymmetric Quantum Mechanics}

The second-order supersymmetric quantum mechanics can be implemented through the iteration of two
first-order transformations. However, by means of this approach it is not revealed in full the
rich set of spectral manipulations which is offered by the second-order technique. This is avoided
if we address the method directly, by assuming that the intertwining operator $B_2^+$ is of
second-order in the derivatives, namely \cite{fe97,fgn98, ro98b,fr06}
\begin{eqnarray}
&&  \hskip1.7cm H_2 B_2^+ = B_2^+ H_0 , \\
&&  \hskip0.2cm H_i=- \frac12 \frac{d^2}{dx^2} + V_i(x), \quad i=0,2 , \\
&&  B_2^+ = \frac12\left(\frac{d^2}{dx^2} - \eta(x)\frac{d}{dx} +
\gamma(x)\right) .
\end{eqnarray}
A calculation similar as for the first-order case leads to the following set of equations
\cite{ff05}:
\begin{eqnarray}
&& \hskip1.5cm V_2 = V_0 - \eta',  \label{v2sususy} \\
&& \frac{\eta''}{2}-\gamma\,' - \eta V_2 = 2V_0' - \eta V_0 , \\
&& \gamma V_2 - \frac{\gamma\,''}{2} = V_0'' - \eta
V_0' + \gamma V_0 .
\end{eqnarray}
By decoupling this system, assuming that $V_0$ is a given solvable potential, we arrive at:
\begin{eqnarray}
&& \hskip3cm \gamma = \frac{\eta'}{2}+\frac{\eta^2}{2} - 2V_0 + d,  \\
 && \frac{\eta\eta''}{2}-\frac{\eta'^2}{4}+\frac{\eta^4}{4}
+ \eta^2\eta' - 2V_0\eta^2 + d\eta^2 + c = 0, \quad c,d\in{\mathbb R}. \label{nlsecond}
\end{eqnarray}
Notice that $V_2$ and $\gamma$ become completely determined once it is found $\eta$; this task is
performed by assuming the ans\"atz
\begin{eqnarray}
& \eta' = -\eta^2 + 2\beta \eta  + 2\xi, \label{anzats}
\end{eqnarray}
where the functions $\beta$ and $\xi$ are to be determined. Therefore
\begin{eqnarray}
& \xi^2{\equiv} c, \quad \beta' + \beta^2 = 2[V_0 - \epsilon], \quad \epsilon=(d+\xi)/2,
\end{eqnarray}
which is once again a Riccati equation for $\beta$. By assuming that $\beta = {u^{(0)}}'/u^{(0)}$
it turns out that $u^{(0)}$ satisfies the stationary Schr\"odinger equation
\begin{eqnarray}
&&  -\frac12 \ {u^{(0)}}'' + V_0  u^{(0)} = \epsilon u^{(0)}.
\end{eqnarray}
Since $\xi = \pm\sqrt{c}$, we get in general two different values for $\epsilon$, $\epsilon_1
\equiv (d +\sqrt{c})/2$, $\epsilon_2 \equiv (d - \sqrt{c})/2$. We thus arrive at a natural
classification of the second-order SUSY transformations, depending on the sign taken by the
parameter $c$ \cite{fr06}.

\subsection{(i) Real case with $c>0$}

For $c>0$ we have that $\epsilon_1, \epsilon_2\in{\mathbb R}$, $\epsilon_1 \neq \epsilon_2$;
denoting by $\beta_1$, $\beta_2$ the corresponding solutions of the Riccati equation it turns out
that (compare Eq.(\ref{anzats}))
\begin{eqnarray}
& \eta' = -\eta^2 + 2 \beta_1 \eta + 2(\epsilon_1-\epsilon_2) ,
\\ & \eta' = -\eta^2 + 2 \beta_2 \eta + 2(\epsilon_2-\epsilon_1) .
\end{eqnarray}
By making the difference of both equations it is obtained:
\begin{eqnarray}
&& \eta =  -\frac{2(\epsilon_1 - \epsilon_2)}{\beta_1 -\beta_2} =
\frac{2(\epsilon_1-\epsilon_2)u^{(0)}_1u^{(0)}_2}{W(u^{(0)}_1,u^{(0)}_2)}
= \frac{W'(u^{(0)}_1,u^{(0)}_2)}{W(u^{(0)}_1,u^{(0)}_2)}. \label{rcw2}
\end{eqnarray}
Notice that now the new potential $V_2$ has no extra singularities with respect to $V_0$ if
$W(u^{(0)}_1,u^{(0)}_2)$ is nodeless (compare Eq.(\ref{v2sususy})).

The spectrum of $H_2$ depends on the normalizability of the two eigenfunctions
$\psi^{(2)}_{\epsilon_{1,2}}$ of $H_2$ associated to $\epsilon_{1,2}$ which belong as well to the
kernel of $B_2$,
\begin{eqnarray}
&& B_2\psi^{(2)}_{\epsilon_j} = 0, \quad H_2 \psi^{(2)}_{\epsilon_j} = \epsilon_j
\psi^{(2)}_{\epsilon_j}, \quad  j=1,2 .
\end{eqnarray}
Their explicit expressions in terms of $u^{(0)}_1$ and $u^{(0)}_2$ become \cite{ff05}
\begin{eqnarray}
&& \psi^{(2)}_{\epsilon_1} \propto
\frac{u^{(0)}_2}{W(u^{(0)}_1,u^{(0)}_2)},
\quad
\psi^{(2)}_{\epsilon_2} \propto
\frac{u^{(0)}_1}{W(u^{(0)}_1,u^{(0)}_2)}.
\end{eqnarray}
We have observed several possible spectral modifications induced on the P\"oschl-Teller
potentials.

\medskip

\noindent(a) Deleting 2 neighbor levels. By taking as seeds two physical eigenfunctions of $H_0$
associated to a pair of neighbor eigenvalues, namely $\epsilon_1 = E_i,$ $\epsilon_2 = E_{i-1}$,
$u_1^{(0)} = \psi_i^{(0)}, \ u_2^{(0)} = \psi_{i-1}^{(0)}$ it is straightforward to show that
\begin{eqnarray}
&& W(u_1,u_2) \propto \sin^{2\lambda + 1}(x) \cos^{2\nu + 1}(x) \,
{\cal W},
\end{eqnarray}
where
\begin{eqnarray}
&& \hskip-0.8cm  {\cal W} = \frac{W\{{}_2F_1[-i,i + \mu; \lambda +
\frac12;\sin^2(x)],{}_2F_1[-i+1,i-1 + \mu; \lambda +
\frac12;\sin^2(x)]\}}{\sin(x) \cos(x)}
\end{eqnarray}
is a nodeless bounded funcion in $[0,\pi/2]$. Hence,
\begin{eqnarray}
V_2 = {(\lambda + 1)(\lambda + 2)\over 2\sin^2(x)} + {(\nu +
1)(\nu + 2)\over 2\cos^2(x)} - (\ln{\cal W})'', \quad \lambda,\nu >1 .
\end{eqnarray}

Notice that the two mathematical eigenfunctions $\psi_{\epsilon_1}^{(2)} , \
\psi_{\epsilon_2}^{(2)}$ of $H_2$ associated to $\epsilon_1 = E_i, \ \epsilon_2 = E_{i-1}$ do not
vanish at $0,\pi/2$,
\begin{eqnarray}
&& \lim_{x\rightarrow 0,\frac{\pi}2}\vert
\psi_{\epsilon_{1,2}}^{(2)}\vert = \infty .
\end{eqnarray}
Thus, $E_{i-1}, E_i\not\in$ Sp$(H_2) = \{E_0, \dots E_{i-2}, E_{i+1},\dots\}$, i.e., we have
deleted the levels $E_{i-1},E_i$ of $H_0$ in order to generate $V_2(x)$.

\medskip

\noindent (b) Creating two new levels. Let us choose now $E_{i-1}<
\epsilon_2<\epsilon_1<E_i,i=0,1,2,\dots$, $E_{-1} \equiv -\infty$, with the corresponding seeds
$u^{(0)}$ as given in Eq.(\ref{vptgs}) with $B_{1,2}=1, \ A_{1,2} = - b_{1,2}/a_{1,2} + q_{1,2}$,
$q_2<0$, $q_1>0$. This choice ensures that $u_2^{(0)}$ and $u_1^{(0)}$ have $i+1$ and $i$
alternating nodes, which produces a nodeless Wronskian \cite{ff05,fmrs02b,sa99}. The divergence of
the $u^{(0)}$solutions for $x\rightarrow 0,\pi/2$ is isolated by expressing
\begin{eqnarray}
&& u_{1,2}^{(0)} = \sin^{1-\lambda}(x) \cos^{1-\nu}(x) {\rm v}_{1,2},
\end{eqnarray}
${\rm v}_{1,2}$ being bounded for $x\in[0,\pi/2]$, ${\rm v}_{1,2}(0)\neq 0 \neq {\rm
v}_{1,2}(\pi/2)$. Since the second term of the Taylor series expansion of ${\rm v}_{1,2}$ is
proportional to $\sin^2(x)$ $\Rightarrow$ ${\rm v}_{1,2}'$ tends to zero as $\sin(x)$ for
$x\rightarrow 0$ and as $\cos(x)$ for $x\rightarrow \pi/2$. Thus,
\begin{eqnarray}
&& W(u_1^{(0)},u_2^{(0)}) = \sin^{3 - 2\lambda}(x) \cos^{3 - 2\nu}(x)
\, {\cal W},
\end{eqnarray}
where ${\cal W} = W({\rm v}_1,{\rm v}_2)/[\sin(x) \cos(x)]$ is nodeless bounded in $[0,\pi/2]$.
Hence
\begin{eqnarray}
&& V_2 = {(\lambda - 3)(\lambda - 2)\over 2\sin^2(x)} + {(\nu -
3)(\nu - 2)\over 2\cos^2(x)} - (\ln{\cal W})'', \quad \lambda, \nu > 3 .
\end{eqnarray}
Since $\lim_{x\rightarrow 0,\frac{\pi}2} \psi_{\epsilon_{1,2}}^{(2)} = 0$ $\Rightarrow$ Sp$(H_2) =
\{E_0, \dots, E_{i-1}, \epsilon_2,\epsilon_1,E_i,\dots\}$, i.e., we have created two new levels
$\epsilon_1,\epsilon_2$ between the originally neighbor ones $E_{i-1}, E_{i}$ for generating
$V_2(x)$.

\medskip

It is worth to mention some other spectral modifications which can be straightforwardly
implemented \cite{cf08}:

\begin{itemize}

\item Isospectral transformations

\item Creating a new level

\item Moving an arbitrary physical level

\item Deleting an original physical level

\end{itemize}

An illustration of the potentials $V_2(x)$ induced by some of the previously discussed spectral
modifications is given in Figure 2.

\begin{figure}[ht]
\centering \epsfig{file=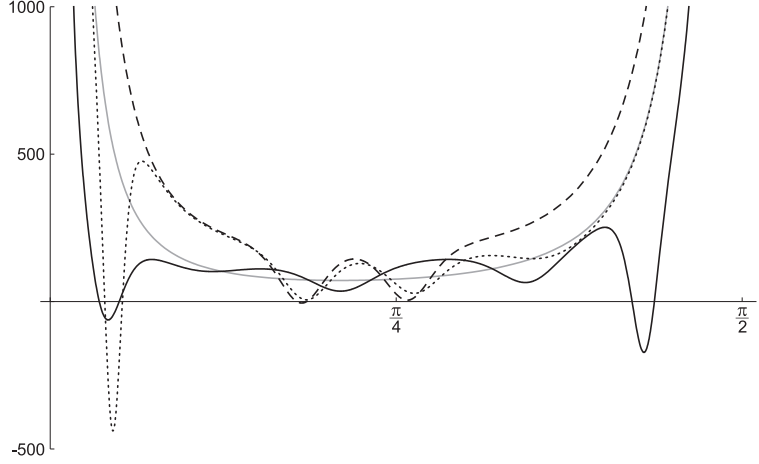, width=10cm}
\caption{Trigonometric P\"oschl-Teller potential $V_0(x)$ for $\lambda = 5, \nu = 8$ (gray curve)
and some second-order SUSY partners $V_2(x)$ in the real case generated by producing the following
spectral modifications: (a) deleting the levels $E_2 = 144.5,$ $E_3= 180.5$ (dashed curve); (b)
creating two new levels at $\epsilon_1 = 128, \ \epsilon_2 = 115.52$ (continuous curve); (e)
moving the level $E_2= 144.5$ up to $\epsilon_1 = 169.28$ (dotted curve) \cite{cf08}}
\end{figure}

\subsection{(ii) Confluent case with $c=0$}

For $\xi=0$ we have that $\epsilon \equiv\epsilon_1= \epsilon_2\in{\mathbb R}$. It this case, once
the solution $\beta$ to the Riccati equation for $\epsilon$ is found, one has to solve the
Bernoulli equation resultant for $\eta$,
\begin{eqnarray}
&&  \hskip2cm \eta' = - \eta^2 + 2 \beta \eta \quad \Rightarrow \quad
\eta = \frac{w'(x)}{w(x)}, \\
&& w(x) = \tilde w_0 +\int \exp[2\int\beta(x)dx]dx
= w_0 + \int_{x_0}^x [u^{(0)}(y)]^2\,dy , \\
&&  \hskip3cm u^{(0)}\propto\exp[\int\beta(x)\,dx] ,
\end{eqnarray}
with $w_0\in {\mathbb R}$. Notice that now $w(x)$ has to be nodeless, which is obtained for seed
solutions $u^{(0)}$ such that
\begin{eqnarray}
&& \lim\limits_{x\rightarrow x_r} u^{(0)} = 0, \quad
\int_{x_0}^{x_r} [u^{(0)}(y)]^2\,dy <\infty \quad {\rm or} \\
&& \lim\limits_{x\rightarrow x_l} u^{(0)} = 0, \quad
\int_{x_l}^{x_0} [u^{(0)}(y)]^2\,dy <\infty,
\end{eqnarray}
and adjusting as well $w_0$, where $[x_l,x_r]$ is the $x$-domain of the involved problem, $x_0\in
[x_l,x_r]$. There is one eigenfunction, $\psi^{(2)}_{\epsilon} \propto u^{(0)}/w$, of $H_2$
associated to $\epsilon$ belonging as well to the kernel of $B_2$, which implies that ${\rm
Sp}(H_2)$ depends of the normalizability of $\psi^{(2)}_{\epsilon}$.

For the P\"oschl-Teller potentials, several possibilities for modifying the initial spectrum have
been observed.

\medskip

\noindent (a) Creating a new level. Let us choose ${\mathbb R}\ni\epsilon\neq E_i$ and $u^{(0)}$
as given in Eq.(\ref{vptgs}) with $A=1, \ B=0$:
\begin{eqnarray}
u^{(0)} & = & \sin^\lambda(x) \cos^\nu(x) \,
{}_2F_1\left(\frac{\mu}{2}  +
\sqrt{\frac{\epsilon}{2}},\frac{\mu}{2}  -
\sqrt{\frac{\epsilon}{2}}; \lambda  +  \frac12;\sin^2(x)\right)  \\
& = & \sin^\lambda(x)\cos^{1 -  \nu}(x) {\rm v}(x),
\end{eqnarray}
where ${\rm v}$ is bounded for $x\in[0,\pi/2]$, ${\rm v}(0)\neq 0 \neq {\rm v}(\pi/2)$. By
calculating the involved integral with $x_0=0$ it is obtained:
\begin{eqnarray}
&& \hskip-1cm w = w_0 + \sum\limits_{m=0}^\infty
\frac{(\frac{\mu}{2} + \sqrt{\frac{\epsilon}{2}})_m(\frac{\mu}{2} -
\sqrt{\frac{\epsilon}{2}})_m \sin^{2\lambda+2m+1}(x)}{(\lambda +
\frac12)_m \, m!(2\lambda + 2m + 1)}
 \nonumber \\
&& \hskip-0.5cm  \times  {}_3F_2\left(\frac{1 \! + \! \lambda \! - \!
\nu }{2} \! - \! \sqrt{\frac{\epsilon}{2}},\frac{1 \! + \! \lambda
\! - \! \nu }{2} \! + \! \sqrt{\frac{\epsilon}{2}},\lambda \! + \! m
\! + \! \frac12; \lambda \! + \! \frac12,\lambda \! + \! m \! + \!
\frac32;\sin^2(x)\right). \label{wconfluente}
\end{eqnarray}
Notice that $w$ is nodeless in $[0,\pi/2]$ for $w_0>0$ and it has one node for $w_0<0$. By
choosing now the $w$ of Eq.(\ref{wconfluente}) with $w_0>0$, we have to isolate anyways a
divergence of kind $\cos^{3-2\nu}(x)$ arising for $x\rightarrow \pi/2$ through the factorization
\begin{eqnarray}
&& w = \cos^{3-2\nu}(x) \, {\cal W},
\end{eqnarray}
${\cal W}$ being a nodeless bounded function for $x\in[0,\pi/2]$. From Eq.(\ref{v2sususy}) we
finally get
\begin{eqnarray}
&& V_2 = {(\lambda - 1)\lambda \over 2\sin^2(x)} + {(\nu-3)(\nu -
2)\over 2\cos^2(x)} - (\ln{\cal W})'', \quad \lambda > 1, \nu > 3 .
\end{eqnarray}
Since $\lim_{x\rightarrow 0,\frac{\pi}2}\psi_{\epsilon}^{(2)} = 0$, it turns out that Sp$(H_2) =
\{ \epsilon, E_n, n=0,1,\dots\}$. It is important to remark that the procedure can be implemented
for solutions associated to $\epsilon > E_0$, i.e., by means of the confluent second-order
transformation one is able to modify the levels above the ground state energy of $H_0$

\medskip

Some additional possibilities of spectral manipulation are now just pointed out \cite{cf08}.

\begin{itemize}

\item Isospectral transformations

\item Deletion of an arbitrary level

\end{itemize}

The potentials $V_2(x)$ induced by some of the mentioned spectral modifications are illustrated in
Figure 3.

\begin{figure}[ht]
\centering \epsfig{file=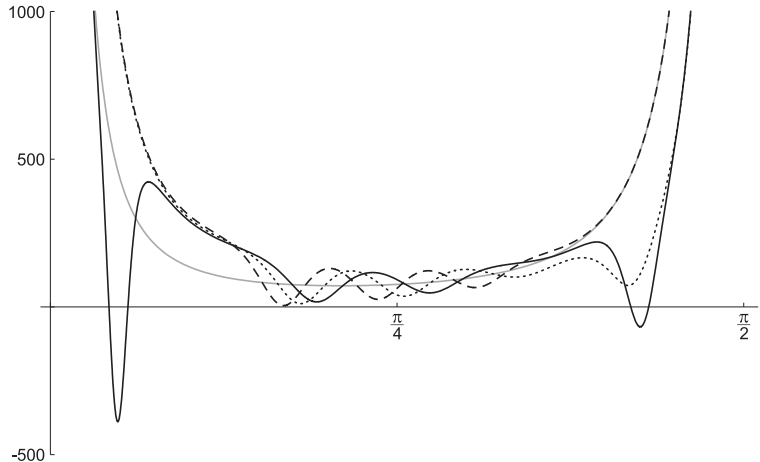, width=10cm}
\caption{Trigonometric P\"oschl-Teller potential $V_0(x)$ for $\lambda = 5, \nu = 8$ (gray curve)
and its second-order SUSY partners $V_2(x)$ in the confluent case, generated through the following
spectral modifications: (a) creating a new level at $\epsilon =147.92$ (continuous curve); (b)
isospectral transformation with $\epsilon = 162$ (dotted curve); (c) deleting the level $E_3 =
180.5$ (dashed curve) \cite{cf08}}
\end{figure}

\subsection{(iii) Complex case with $c<0$}

For $c<0$ it happens that $\epsilon_{1,2}\in{\mathbb C}$, $\epsilon_2 = \bar\epsilon_1$. As we
want that $V_2$ to be real, it has to be chosen $\beta_2 = \bar\beta_1$. Thus,
\begin{eqnarray}
&& \eta = -\frac{2{\rm Im}(\epsilon_1)}{{\rm Im}[\beta_1]} =
\frac{w'}{w}, \quad w = \frac{W(u^{(0)}_1,\bar
u^{(0)}_1)}{2(\epsilon_1 - \bar \epsilon_1)},
\end{eqnarray}
where $u^{(0)}_1$ is a complex solution of the Schr\"odinger equation associated to
$\epsilon_{1}\in{\mathbb C}$. As in the previous cases, $w$ has to be nodeless in order to avoid
the arising of extra singularities in $V_2$. For this to happens, it is sufficient that
\begin{eqnarray}
&& \lim_{x\rightarrow x_r} u^{(0)}_1 = 0 \quad {\rm or} \quad
\lim_{x\rightarrow x_l} u^{(0)}_1 = 0 .
\end{eqnarray}
In both cases we get that ${\rm Sp}(H_2) = {\rm Sp}(H_0)$.

We illustrate now the technique through the trigonometric P\"oschl-Teller potentials, for which we
take $\epsilon\in{\mathbb C}$, $u^{(0)}$ as in Eq.(\ref{vptgs}) with $A=1, \ B=0$ $\Rightarrow$
$u^{(0)}(0) = 0$, $\vert u^{(0)}\vert \rightarrow \infty$ when $x\rightarrow \pi/2$. It is
appropriate to express now
\begin{eqnarray}
&& u^{(0)} = \sin^{\lambda}(x) \cos^{1-\nu}(x) {\rm v} .
\end{eqnarray}
Hence, it is obtained:
\begin{eqnarray}
&&  \hskip2.5cm w = \sin^{2\lambda + 1}(x) \cos^{3 -
2\nu}(x) \, {\cal W}, \\
&&  \hskip2.6cm {\cal W} = \frac {W({\rm v},\bar {\rm v})}{2(\epsilon - \bar \epsilon)\sin(x)
\cos(x)}, \\
&& V_2 = {(\lambda + 1)(\lambda + 2)\over 2\sin^2(x)} + {(\nu -
3)(\nu - 2)\over 2\cos^2(x)} - (\ln{\cal W})'', \quad \lambda > 1,
\nu > 3 . \label{v2ptcom}
\end{eqnarray}
The potentials $V_0$ and $V_2$ turn out to be isospectral. An illustration of $V_0(x)$ and the
$V_2(x)$ of Eq.(\ref{v2ptcom}) is given in Figure 4.

A general conclusion, which can be inferred of the previously discussed first and second-order
SUSY QM but it is valid for an arbitrary order, is that the spectra of the initial and new
Hamiltonians differ little. Thus, one would expect that some properties which are somehow related
to the corresponding spectral problems would be essentially the same. We will see next that this
is true, in particular, for the algebraic structure ruling the two involved Hamiltonians.

\begin{figure}[ht]
\centering \epsfig{file=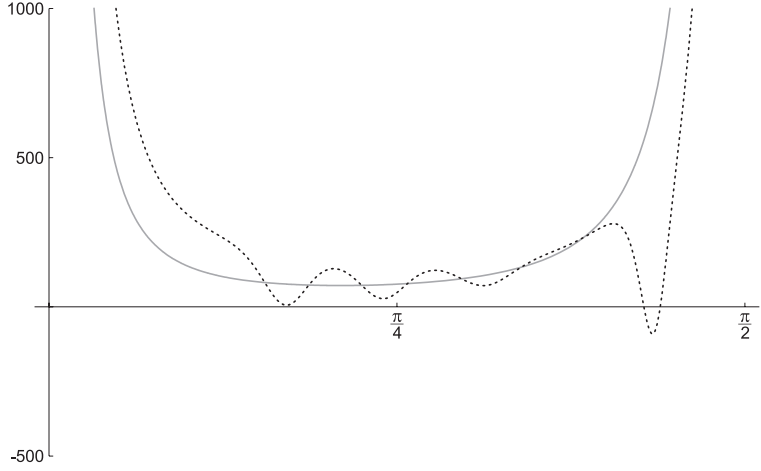, width=10cm}
\caption{Trigonometric P\"oschl-Teller potential $V_0(x)$ for $\lambda = 5, \nu = 8$ (gray curve)
and its second-order SUSY partner $V_2(x)$ in the complex case generated by using $\epsilon =
176.344 + 1.5i$ and a solution $u^{(0)}$ vanishing at the origin (dotted curve) \cite{cf08}}
\end{figure}

\section{Algebra of the SUSY partner Hamiltonians}

Given an initial Hamiltonian $H_0$, which is characterized by a certain algebraic structure, it is
natural to pose the following questions:

\begin{itemize}

\item What is the algebra ruling the potentials generated through SUSY QM?

\item What other properties of $H_0$ are inherited by its SUSY partner Hamiltonians?

\end{itemize}

These questions will be analyzed here through a simple system which algebraic structure
generalizes the standard Heisenberg-Weyl (oscillator) algebra. Thus, let us consider a
one-dimensional Schr\"odinger Hamiltonian
\begin{eqnarray}
&& H_0 = -\frac12 \frac{d^2}{dx^2} + V_0(x),
\end{eqnarray}
whose eigenvectors $\vert\psi_n^{(0)}\rangle$ and eigenvalues $E_n$ satisfy (for calculational
convenience we switch now to the Dirac bra and ket notation)
\begin{eqnarray}
&&  \hskip0.1cm H_0 \vert\psi_n^{(0)}\rangle = E_n \vert\psi_n^{(0)}\rangle,
\quad E_0<E_1<E_2< \dots \\
&& \langle \psi_m^{(0)}\vert \psi_n^{(0)}\rangle = \delta_{mn},
\qquad \sum\limits_{m=0}^{\infty} \vert\psi_m^{(0)}
\rangle\langle\psi_m^{(0)}\vert = 1 .
\end{eqnarray}
It is supposed that there is an analytic dependence between the eigenvalues $E_n$ and the index
labeling them, namely,
\begin{eqnarray}
E_n\equiv E(n) .
\end{eqnarray}

Let us define now the {\it intrinsic algebra} of $H_0$, by introducing a pair of annihilation and
creation operators $a_0^\pm$ whose action onto the eigenvector of $H_0$ is given by
\begin{eqnarray}
&& a_0^- \vert\psi_n^{(0)}\rangle = r_{I}(n)
\vert\psi_{n-1}^{(0)}\rangle, \quad a_0^+ \vert\psi_n^{(0)}\rangle =
\bar r_{I}(n+1)
\vert\psi_{n+1}^{(0)}\rangle, \\
&&  \hskip0.8cm r_{I}(n) = e^{i\alpha[E(n)-E(n-1)]} \ \sqrt{E(n) - E_0},
\quad \alpha \in {\mathbb R} .
\end{eqnarray}
With these definitions it is straightforward to show that:
\begin{eqnarray}
&& a_0^+ a_0^- = H_0 - E_0 .
\end{eqnarray}
It is introduced the number operator $N_0$ through its action onto the eigenvectors of $H_0$:
\begin{eqnarray}
&& N_0 \vert\psi_{n}^{(0)}\rangle = n\vert\psi_{n}^{(0)}\rangle, \quad
[N_0,a_0^\pm] = \pm a_0^\pm .
\end{eqnarray}

Now, the intrinsic algebra of the systems is characterized by
\begin{eqnarray}
&& a_0^+ a_0^- = E(N_0) - E_0, \quad a_0^- a_0^+ =
E(N_0+1) - E_0 , \\
&&  \hskip1cm [a_0^-,a_0^+] = E(N_0+1) - E(N_0) \equiv f(N_0) .
\end{eqnarray}
It is important to realize that for general $E(n)$ the intrinsic algebra becomes nonlinear (since
$f(N_0)$ is not necessarily a linear function of $N_0$). However, this intrinsic nonlinear algebra
can be linearized by deforming the corresponding annihilation and creation operators in the way
\begin{eqnarray}
&& a_{0_{L}}^- = \frac{r_{L}(N_0 + 1)}{r_{I}(N_0 + 1)} \, a_0^-,
\quad a_{0_{L}}^+ = \frac{r_{L}(N_0)}{r_{I}(N_0)} \, a_0^+ , \quad
r_{L}(n) = e^{i\alpha f(n-1)} \, \sqrt{n}, \\
&&  \hskip-0.3cm [N_0 , a_{0_{L}}^\pm] = \pm a_{0_{L}}^\pm, \quad a_{0_{L}}^+ a_{0_{L}}^- = N_0,
\quad a_{0_{L}}^- a_{0_{L}}^+ = N_0 + 1, \quad
[a_{0_{L}}^-, a_{0_{L}}^+] = 1 .
\end{eqnarray}

In particular, for the harmonic oscillator Hamiltonian the intrinsic algebraic structure reduces
to the standard Heisenberg-Weyl algebra:
\begin{eqnarray}
&& E(n) = n + \frac12, \quad
f(n) = E(n+1) - E(n) = 1, \\
&&  \hskip1.5cm [N_0,a_0^\pm] = \pm a_0^\pm , \quad [a_0^-,a_0^+] = 1 .
\end{eqnarray}
On the other hand, for the trigonometric P\"oschl-Teller potentials we recover the su(1,1)
algebra:
\begin{eqnarray}
&& E(n) =  \frac{(\mu + 2n)^2}2 , \quad
f(n) = E(n+1) - E(n) = 4n + 2\mu + 2, \\
&&  \hskip1.8cm [N_0,a_0^\pm] = \pm a_0^\pm , \quad [a_0^-,a_0^+] = 4 N_0 + 2\mu + 2 .
\end{eqnarray}

Let us characterize now the algebraic structure of the SUSY partner Hamiltonians of $H_0$
\cite{fhr07}. In order to illustrate the procedure, we restrict ourselves just to a second-order
SUSY transformation which creates two new levels at the positions $\epsilon_1, \ \epsilon_2$ for
$H_2$, namely,
\begin{eqnarray}
&&  {\rm Sp}(H_2) = \{\epsilon_1, \epsilon_2,E_n, n=0,1,\dots \}
\end{eqnarray}
Notice that the eigenstates of $H_2$, $\vert\psi_n^{(2)} \rangle$, associated to $E_n,
n=0,1,\dots$ are obtained from $\vert \psi_n^{(0)} \rangle$ and vice versa through the action of
the intertwining operators $B_2^+$, $B_2$ in the way:
\begin{eqnarray}
&& \vert\psi_n^{(2)}\rangle =  \frac{B_2^+\vert
\psi_n^{(0)}\rangle}{\sqrt{(E_n-\epsilon_1)(E_n-\epsilon_2)}},
\qquad \vert \psi_n^{(0)}\rangle  =  \frac{B_2\vert
\psi_n^{(2)}\rangle}{\sqrt{(E_n-\epsilon_1)(E_n-\epsilon_2)}}.
\label{psi}
\end{eqnarray}
The eigenstates of $H_2$ constitute a complete orthonormal set,
\begin{eqnarray}
&&  \hskip2cm H_2 \vert \psi_n^{(2)} \rangle = E_n \vert \psi_n^{(2)} \rangle,
\quad H_2 \vert \psi_{\epsilon_i}^{(2)}\rangle = \epsilon_i \vert
\psi_{\epsilon_i}^{(2)}\rangle ,  \\
&&  \hskip-1.5cm \langle \psi_{\epsilon_i}^{(2)} \vert \psi_n^{(2)} \rangle = 0,
\ \langle \psi_m^{(2)} \vert  \psi_n^{(2)} \rangle =
\delta_{mn}, \ \langle \psi_{\epsilon_i}^{(2)}
\vert\psi_{\epsilon_j}^{(2)} \rangle =
\delta_{ij} ,  \ n,m=0,1,\dots, \ i,j=1,2,
\\ &&  \hskip2.5cm \sum\limits_{l=1}^{2} \vert \psi_{\epsilon_l}^{(2)}\rangle \langle
\psi_{\epsilon_l}^{(2)} \vert + \sum\limits_{m = 0}^{\infty} \vert
\psi_m^{(2)} \rangle\langle \psi_m^{(2)} \vert = 1 .
\end{eqnarray}

It is important to notice that the two new levels $\epsilon_1, \epsilon_2$ can be placed at
positions essentially arbitrary (either both below the initial ground state $E_0$ or in between
two neighbor physical levels $E_{i-1}, E_i$). This fact somehow breaks the symmetry defined by
$E(n)$. Thus, it should be clear that we need to isolate, in a sense, the two new levels
$\epsilon_1, \epsilon_2$. With this aim, it is defined now the number operator $N_2$ through the
following action onto the eigenstates of $H_2$:
\begin{eqnarray}
&& N_2 \vert \psi_n^{(2)} \rangle = n \vert \psi_n^{(2)} \rangle, \ \ N_2
\vert \psi_{\epsilon_i}^{(2)}\rangle = 0, \ \ n=0,1,\dots \ i=1,2 .
\end{eqnarray}
Departing now from the operators $a_0^\pm$ of the intrinsic algebra of $H_0$, let us construct
those of the {\it natural algebra} of $H_2$ \cite{fhr07}, which represent a generalization of the
operators introduced previously for the SUSY partners of the harmonic oscillator
\cite{mi84,fhn94,fnr95,fh99},
\begin{eqnarray}
&& a^\pm_{2_{N}} = B_2^+ a_0^\pm B_2 .
\end{eqnarray}
An schematic representation of these operators is given in Figure 5. Their action onto the
eigenvectors of $H_2$ is given by

\begin{figure}[ht]
\centering \epsfig{file=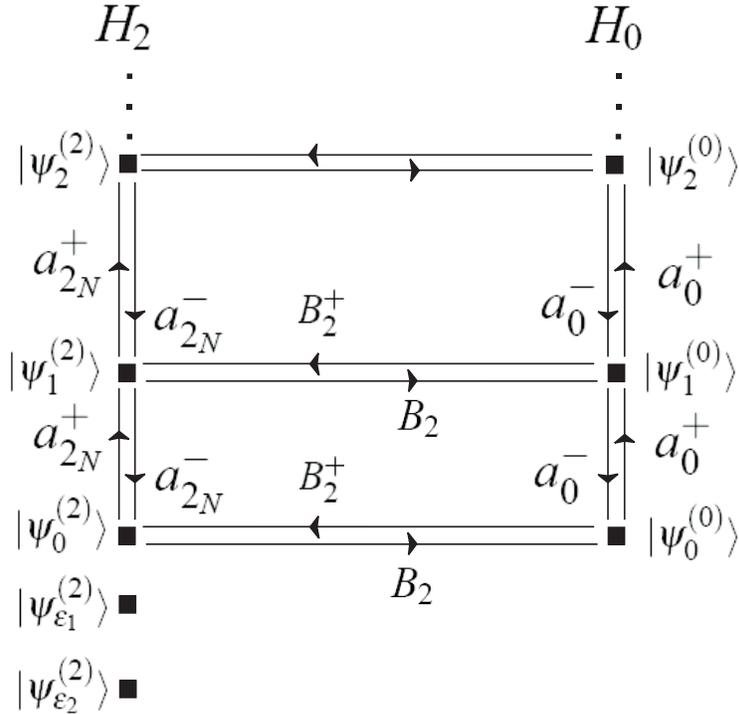, width=10cm}
\caption{Representation of the annihilation and creation operators $a^\pm_{2_{N}}$ of the natural
algebra of $H_2$.}
\end{figure}

\begin{eqnarray}
&&  \hskip-0.8cm a^\pm_{2_{N}} \vert \psi_{\epsilon_i}^{(2)}\rangle = 0 ,  \quad
a^-_{2_{N}} \vert \psi_n^{(2)} \rangle = r_{N}(n) \
\vert \psi_{n-1}^{(2)} \rangle, \quad  a^+_{2_{N}} \vert
\psi_n^{(2)} \rangle = \bar r_{N}(n + 1) \ \vert \psi_{n+1}^{(2)}
\rangle,  \\
&& \hskip-0.1cm r_{N}(n) = \left\{\prod\limits_{i=1}^2[E(n) -
\epsilon_i][E(n-1)-\epsilon_i] \right\}^\frac12 r_{I}(n), \quad i=1,2 , \ n=0,1,\dots.
\end{eqnarray}
Using these equations it is found that
\begin{eqnarray}
&& [a^-_{2_{N}} \! , \! a^+_{2_{N}}\!] \! = \!
\left[\bar r_{N}(\! N_2 \!\! + \!\! 1) r_{N}(\! N_2 \!\! + \!\! 1) \! - \! \bar
r_{N}(N_2) r_{N}(N_2)\right] \!\! \sum\limits_{m=0}^{\infty} \!\! \vert\psi_{m}^{(2)}
\rangle \! \langle\psi_{m}^{(2)} \vert .
\end{eqnarray}

Let us build up now the annihilation and creation operators $a_{2}^\pm$ which generate the
intrinsic algebra of $H_2$,
\begin{eqnarray}
&& \hskip-1.2cm a_2^- = \frac{r_{I}(N_2 +
1)}{r_{N}(N_2 + 1)} \, a_{2_{N}}^-, \quad a_2^+ =
\frac{r_{I}(N_2)}{r_{N}(N_2)} \, a_{2_{N}}^+ , \quad
[a_2^-,a_2^+] = f(N_2) \sum\limits_{m=0}^\infty \vert\psi_{m}^{(2)}
\rangle\langle\psi_{m}^{(2)}\vert .
\end{eqnarray}
Their action onto the eigenstates of $H_2$ reads
\begin{eqnarray}
&& \hskip0.9cm a_2^\pm \vert \psi_{\epsilon_i}^{(2)} \rangle = 0, \quad
a_2^- \vert \psi_n^{(2)} \rangle = r_{I}(n) \vert \psi_{n-1}^{(2)}
\rangle , \\
&& a_2^+ \vert \psi_n^{(2)} \rangle = \bar r_{I}(n +
1)\vert \psi_{n+1}^{(2)} \rangle, \quad i=1,2, \ n=0,1,\dots
\end{eqnarray}
Notice that this algebra of $H_2$ is the same as the intrinsic algebra of $H_0$ when we work on
the restriction to the subspace spanned by the eigenvectors $\vert \psi_n^{(2)} \rangle$
associated to the isospectral part of the spectrum.

Let us construct, finally, the annihilation and creation operators $a_{2_{L}}^\pm$ of the linear
algebra of $H_2$,
\begin{eqnarray}
&& \hskip0.8cm a_{2_{L}}^- = \frac{r_{L}(N_2 + 1)}{r_{I}(N_2 + 1)} \, a_{2}^- ,
\quad a_{2_{L}}^+ =
\frac{r_{L}(N_2)}{r_{I}(N_2)} \, a_{2}^+,  \\
&& [N_2, a_{2_{L}}^\pm] =  \pm a_{2_{L}}^\pm ,
\qquad [a_{2_{L}}^-, a_{2_{L}}^+] =
\sum\limits_{m=0}^\infty \vert \psi_m ^{(2)}\rangle\langle \psi_m^{(2)} \vert .
\end{eqnarray}
Their action onto the eigenstates of $H_2$ is given by
\begin{eqnarray}
&&  \hskip0.8cm a_{2_{L}}^\pm \vert \psi_{\epsilon_i}^{(2)} \rangle = 0, \quad
a_{2_{L}}^- \vert \psi_n^{(2)} \rangle = r_{L}(n) \vert \psi_{n-1}^{(2)}
\rangle , \\
&&  a_{2_{L}}^+ \vert \psi_n^{(2)} \rangle = \bar r_{L}(n +
1)\vert \psi_{n+1}^{(2)} \rangle,   \quad i=1,2, \ n=0,1,\dots
\end{eqnarray}
Once again, this algebra of $H_2$ coincides with the corresponding one of $H_0$ on the subspace
spanned by $\{ \vert \psi_n^{(2)} \rangle, n=0,1,\dots \}$.

\section{Coherent states}

Once identified the algebraic structures associated to $H_0$ and $H_2$, it would be important to
generate the corresponding coherent states. It is well know that there are several definitions of
coherent states \cite{ks85,pe86,gaa00,fv09}; here we will derive them as eigenstates with complex
eigenvalues of the annihilation operators of the systems. Since there are available several
annihilation operators for $H_0$ and $H_2$, we will call them with the same name as the
corresponding associated algebra.

\subsection{Coherent states of $H_0$}

Since for $H_0$ we identified two different algebraic structures, intrinsic and linear one, we
will look for the two corresponding families of coherent states.

\subsubsection{Intrinsic coherent states $\vert z,\alpha\rangle_0$ of $H_0$}

The intrinsic coherent states $\vert z,\alpha\rangle_0$ of $H_0$ satisfy:
\begin{eqnarray}
&& a_0^-\vert z,\alpha\rangle_0 = z\vert z,\alpha\rangle_0, \quad
z\in{\mathbb C} . \label{dcsi0}
\end{eqnarray}
Expanding $\vert z,\alpha\rangle_0$ in the basis of eigenstates of $H_0$,
\begin{eqnarray}
&& \vert z,\alpha\rangle_0 = \sum\limits_{n=0}^\infty c_n \vert \psi_n^{(0)} \rangle,
\end{eqnarray}
and substituting in Eq.(\ref{dcsi0}), it is obtained a recurrence relationship for $c_n$,
\begin{eqnarray}
&& c_n = \frac{z}{r_{I}(n)}c_{n - 1} =
\frac{z^n}{r_{I}(n)\dots r_{I}(1)}c_0 =
\frac{e^{-i\alpha(E_n-E_0)}z^n}{\sqrt{(E_n-E_0)\dots(E_1-E_0)}}c_0  .
\end{eqnarray}
By choosing $c_0\in {\mathbb R}^+$ so that $\vert z,\alpha\rangle_0$ is normalized we finally get
\begin{eqnarray}
&& \vert z,\alpha\rangle_0 = \left(\sum\limits_{m=0}^{\infty} \frac{\vert
z\vert^{2m}}{\rho_m}\right)^{-\frac12} \sum\limits_{m =
0}^{\infty}e^{-i\alpha (E_m - E_0)}
\frac{z^m}{\sqrt{\rho_m}}\vert \psi_m^{(0)} \rangle , \\
&& \hskip1cm \rho_m =
\bigg\{\begin{matrix}
1 & {\rm for} \ m=0 , \\
(E_m - E_0)\dots(E_1 - E_0) & {\rm for} \ m > 0 .
\end{matrix}
\label{momentsi0}
\end{eqnarray}

An important property that should be satisfied is the so-called completeness relationship, namely,
\begin{eqnarray}
&& \int \vert z,\alpha\rangle_0 \ {}_0\langle z,\alpha \, \vert d\mu(z) = 1, \label{incom}
\end{eqnarray}
where the positive defined measure $d\mu(z)$ can be expressed as
\begin{eqnarray}
&& d\mu(z) = \frac{1}{\pi} \left(\sum\limits_{m=0}^{\infty}\frac{\vert
z\vert^{2m}}{\rho_m}\right) \rho(\vert z\vert^2) \, d^2 z .
\end{eqnarray}
Working in polar coordinates and changing variables $y=\vert z\vert ^2$, it turns out that
$\rho(y)$ must satisfy
\begin{eqnarray}
&& \int_0^\infty y^m \rho(y) \, dy = \rho_m, \quad m = 0,1,\dots \label{mpi0}
\end{eqnarray}
This is a moment problem depending on $\rho_n$, i.e., on the spectral function $E(n)$
\cite{fhn94,fh99,jr99,sps99,sp00,kps01,agmkp01}. If one is able to solve Eq.(\ref{mpi0}) for
$\rho(y)$, the completeness relation becomes valid.

Once Eq.(\ref{incom}) is satisfied, an arbitrary quantum state can be expressed in terms of
coherent states. In particular, an intrinsic coherent state $\vert z',\alpha\rangle_0$ admits the
decomposition,
\begin{eqnarray}
&& \vert z',\alpha\rangle_0 = \int \vert z,\alpha\rangle_0 \ {}_0\langle z,\alpha
\vert z',\alpha\rangle_0 \, d\mu(z) ,
\end{eqnarray}
where the reproducing kernel ${}_0\langle z,\alpha \vert z',\alpha\rangle_0$ is given by
\begin{eqnarray}
&&  {}_0\langle z,\alpha \vert z',\alpha\rangle_0 \! = \! \left(\sum\limits_{m=0}^{\infty}
\frac{\vert z \vert^{2m}}{\rho_m}\right)^{-1/2} \left(\sum\limits_{m=0}^{\infty}
\frac{\vert z'\vert^{2m}}{\rho_m}\right)^{-1/2} \left(\sum\limits_{m=0}^{\infty}
\frac{(\bar zz')^m}{\rho_m}\right).
\end{eqnarray}
As $\vert z=0,\alpha\rangle_0 = \vert\psi_0^{(0)} \rangle$, it turns out that the eigenvalue $z=0$
is non-degenerate. If the system is initially in an intrinsic coherent state $\vert
z,\alpha\rangle_0$, then it will be at any time in an intrinsic coherent state,
\begin{eqnarray}
&& U_0(t) \vert z,\alpha\rangle_0 = e^{-itE_0}\vert z,\alpha+t\rangle_0,
\end{eqnarray}
where $U_0(t) = \exp(-i H_0 t)$ is the evolution operator of the system.

\subsubsection{Linear coherent states $\vert z,\alpha\rangle_{0_{L}}$ of $H_0$}

Now, let us look for the linear coherent states such that
\begin{eqnarray}
&& a_{0_{L}}^- \vert z,\alpha\rangle_{0_{L}} = z
\vert z,\alpha\rangle_{0_{L}}, \quad z\in{\mathbb C}.
\end{eqnarray}
A procedure similar to the previous one leads straightforwardly to
\begin{eqnarray}
&& \vert z,\alpha\rangle_{0_{L}} = e^{-\frac{\vert z
\vert^2}2} \sum\limits_{m = 0}^{\infty} e^{-i\alpha(E_m -
E_{0})}\frac{z^m}{\sqrt{m!}} \, \vert\psi_m^{(0)}\rangle ,
\end{eqnarray}
which, up to the phases, has the form of the standard coherent states. Now the completeness
relationship is automatically satisfied,
\begin{eqnarray}
&& \frac{1}{\pi} \int \vert z,\alpha\rangle_{0_{L}} \
{}_{0_{L}}\langle z, \alpha \vert \, d^2 z = 1 .
\end{eqnarray}
Therefore, an arbitrary linear coherent state $\vert z',\alpha\rangle_{0_{L}}$ is expressed in
terms of linear coherent states,
\begin{eqnarray}
&& \vert z',\alpha\rangle_{0_{L}} = \frac{1}{\pi}\int \vert
z,\alpha\rangle_{0_{L}} \ {}_{0_{L}}\langle z,
\alpha \vert z',\alpha\rangle_{0_{L}} \, d^2 z ,
\end{eqnarray}
where the reproducing kernel is given by
\begin{eqnarray}
&& {}_{0_{L}}\langle z, \alpha \vert
z',\alpha\rangle_{0_{L}} = \exp\left(-\vert z
\vert^2/2 + \bar z z' - \vert z' \vert^2/2 \right) .
\end{eqnarray}
Notice that, once again, the only eigenstate of $H_0$ which is also a linear coherent state is the
ground state, $\vert z=0,\alpha\rangle_{0_{L}} = \vert\psi_0^{(0)}\rangle$. Moreover, since
$[a_{0_{L}}^-,a_{0_{L}}^+]=1$ then the linear coherent states also result from the action of the
displacement operator $D_{L}(z)$ onto the ground state $\vert\psi_0^{(0)}\rangle$:
\begin{eqnarray}
&& \vert z,\alpha\rangle_{0_{L}} = D_{L}(z)\vert\psi_0^{(0)}\rangle = \exp(za_{0_{L}}^+ - \bar z
a_{0_{L}}^-)\vert\psi_0^{(0)}\rangle .
\end{eqnarray}
Finally, an initial linear coherent state evolves in time as a linear coherent state, namely,
\begin{eqnarray}
&& U_0(t) \vert z,\alpha\rangle_{0_L} = e^{-itE_0}\vert z,\alpha+t\rangle_{0_L}.
\end{eqnarray}

\subsection{Coherent states of $H_2$}

Let us remember that we have analyzed three different algebraic structures for $H_2$, namely,
natural, intrinsic and linear ones; thus three families of coherent states will be next
constructed.

\subsubsection{Natural coherent states $\vert z,\alpha\rangle_{2_{N}}$ of $H_2$}

The natural coherent states $\vert z,\alpha\rangle_{2_{N}}$ of $H_2$ are defined by
\begin{eqnarray}
&& a_{2_{N}}^- \vert z,\alpha\rangle_{2_{N}} = z \vert z,\alpha\rangle_{2_{N}}.
\end{eqnarray}
A procedure similar to the previously used leads to
\begin{eqnarray}
&& \vert z,\alpha\rangle_{2_{N}}  = \left[\sum\limits_{m = 0}^{\infty}
\frac{\vert z\vert^{2m}}{\widetilde\rho_m}\!\right]^{-\frac12} \sum\limits_{m
= 0}^{\infty} \! e^{-i\alpha (E_{m } -
E_{0})}\frac{z^{m}}{\sqrt{\widetilde\rho_m}}\vert\psi_{m }^{(2)}\rangle ,
\end{eqnarray}
where
\begin{eqnarray}
&& \hskip-1cm \widetilde\rho_m   = \bigg\{\begin{matrix} 1 & {\rm for} \ m=0,  \\
\rho_{m}  \prod\limits_{i=1}^2
(E_{m}  - \epsilon_i) (E_{m -1}  -
\epsilon_i)^2  \dots \! (E_{1}  - \epsilon_i)^2(E_{0}  - \epsilon_i) & {\rm for}\ m > 0 .
\end{matrix}
\label{momentsn2}
\end{eqnarray}
Notice that now the completeness relationship has to be modified to include the projector onto the
subspace spanned by the eigenvectors of $H_2$ associated to the two isolated levels $\epsilon_1,
\epsilon_2$, namely,
\begin{eqnarray}
&& \sum\limits_{i=1}^2  \vert \psi_{\epsilon_i}^{(2)}\rangle \langle
\psi_{\epsilon_i}^{(2)} \vert  +  \int \vert
z,\alpha\rangle_{2_{N}} \, {}_{2_{N}}\langle
z,\alpha \vert \, d\widetilde\mu(z)  =  1 ,
\end{eqnarray}
where
\begin{eqnarray}
&& d\widetilde\mu(z) = \frac{1}{\pi} \left(\sum\limits_{m = 0}^{\infty}
\frac{\vert z\vert^{2m}}{\widetilde\rho_m}\right)
\widetilde\rho(\vert z\vert^2) \, d^2z, \\
&& \hskip0.8cm \int\limits_0^\infty y^m \widetilde\rho(y) \, dy = \widetilde \rho_m ,
\quad m \geq 0 . \label{mpn}
\end{eqnarray}
It is clear that the moment problem characterized by Eq.(\ref{mpn}) is more complicated than the
one associated to Eq.(\ref{mpi0}) (compare the moments given by Eqs.(\ref{momentsi0}) and
(\ref{momentsn2})). Since $B_2 \vert\psi_{\epsilon_i}^{(2)}\rangle = a^-_{2_{N}} \vert
\psi_{\epsilon_i}^{(2)}\rangle = 0, \ i=1,2$ and $a^-_{2_{N}} \vert \psi_{0}^{(2)}\rangle = 0$,
then the eigenvalue $z=0$ of $a_{2_{N}}^-$ is 3-fold degenerate. Finally, as for the coherent
states of $H_0$, it turns out that the natural coherent states $\vert z,\alpha\rangle_{2_{N}}$
evolve coherently with $t$,
\begin{eqnarray}
&& U_2(t) \vert z,\alpha\rangle_{2_{N}} = e^{-itE_{0}}
\vert z,\alpha + t\rangle_{2_{N}},
\end{eqnarray}
where $U_2(t) = \exp(- i H_2 t)$ is the evolution operator of the system characterized by $H_2$.

\subsubsection{Intrinsic coherent states $\vert z,\alpha\rangle_{2}$ of $H_2$}

The intrinsic coherent states $\vert z,\alpha\rangle_{2}$ of $H_2$ are eigenstates with complex
eigenvalue $z\in{\mathbb C}$ of the intrinsic annihilation operator $a_{2}^-$ of $H_2$. A
treatment similar to the previous one leads to
\begin{eqnarray}
&& \vert z,\alpha\rangle_2 = \left(\sum\limits_{m=0}^{\infty} \frac{\vert
z\vert^{2m}}{\rho_m}\right)^{-\frac12} \sum\limits_{m = 0}^{\infty}
e^{-i\alpha (E_m -
E_0)}\frac{z^m}{\sqrt{\rho_m}}\vert\psi_m^{(2)}\rangle .
\end{eqnarray}
The modified completeness relationship reads now
\begin{eqnarray}
&& \sum\limits_{i=1}^2 \vert \psi_{\epsilon_i}^{(2)}\rangle \langle
\psi_{\epsilon_i}^{(2)} \vert + \int \vert z,\alpha\rangle_2 \,
{}_2\langle z,\alpha \vert \, d\mu(z) = 1 ,
\end{eqnarray}
where $d\mu(z)$ is the same as for $H_0$. Since $a_2^- \vert \psi_{\epsilon_i}^{(2)} \rangle = 0,
\ i=1,2$ and $\vert z=0,\alpha\rangle_2 = \vert \psi_{0}^{(2)}\rangle$, the eigenvalue $z=0$ of
$a_2^-$ turns out to be $3$-fold degenerate. As for the natural coherent states of $H_2$, the
intrinsic coherent states $\vert z,\alpha\rangle_2$ evolve in time in a coherent way.

\subsubsection{Linear coherent states $\vert z,\alpha\rangle_{2_{L}}$ of $H_2$}

Let us consider, finally, the linear coherent states $\vert z,\alpha\rangle_{2_{L}}$ of $H_2$ as
eigenstates of $a_{2_{L}}^-$ with eigenvalue $z\in{\mathbb C}$. A standard procedure produces now
\begin{eqnarray}
&& \vert z,\alpha\rangle_{2_{L}} = e^{-\frac{\vert z
\vert^2}2} \sum\limits_{m = 0}^{\infty} e^{-i\alpha(E_m -
E_{0})}\frac{z^m}{\sqrt{m!}} \, \vert\psi_m^{(2)}\rangle .
\end{eqnarray}
The completeness relationship is expressed as
\begin{eqnarray}
&& \sum\limits_{i=1}^2 \vert \psi_{\epsilon_i}^{(2)}\rangle \langle
\psi_{\epsilon_i}^{(2)} \vert + \frac{1}{\pi}\int \vert
z,\alpha\rangle_{2_{L}} \, {}_{2_{L}}\langle
z,\alpha \vert \, d^2z = 1 .
\end{eqnarray}
Once again, the eigenvalue $z=0$ of $a_{2_{L}}^-$ is $3$-fold degenerate. Moreover, the linear
coherent states $\vert z,\alpha\rangle_{2_{L}}$ appear from the action of the displacement
operator $D_{2_{L}}(z)$ onto the extremal state $\vert \psi_{0}^{(2)}\rangle$:
\begin{eqnarray}
&& \vert z,\alpha\rangle_{2_{L}} = D_{2_{L}}(z)\vert
\psi_{0}^{(2)}\rangle = \exp(za_{2_{L}}^+ -  \bar z
a_{2_{L}}^-)\vert \psi_{0}^{(2)}\rangle .
\end{eqnarray}
As in the previous cases, the linear coherent states of $H_2$ evolve in time coherently.

\bigskip

Let us end up these two sections, about the algebraic structure of the SUSY partner Hamiltonians
and the corresponding coherent states, with the following conclusions.

\begin{itemize}

\item The intrinsic and linear algebraic structures of $H_0$ are inherited by $H_2$ on the
subspace associated to the levels of the isospectral part of the spectrum

\item This reflects as well in the corresponding coherent states of $H_2$, linear and intrinsic
ones, whose expressions on the same subspace are equal to the corresponding ones of $H_0$

\item We generalized successfully the construction of the natural algebra of $H_2$ when the
initial potential $V_0(x)$ is more general than the harmonic oscillator

\end{itemize}

\section{SUSY partners of periodic potentials}

The exactly solvable models are important since the relevant physical information is encoded in a
few expressions. Moreover, they are ideal either to test the accuracy (and convergence) of
numerical techniques, or to implement some analytic approximate methods, as perturbation theory.
Unfortunately, the number of exactly solvable periodic potentials is small, so it is important to
enlarge this class. As we saw in the previous sections, a simple method to do the job is
supersymmetric quantum mechanics. Before applying the technique to periodic potentials, however,
let us review first the way in which it is classified the energy axis in allowed and forbidden
energy bands.

For one-dimensional periodic potentials such that $V_0(x+T)=V_0(x)$, the Schr\"odinger equation
\begin{eqnarray}
&& - \frac12 \psi''(x) + V_0(x) \psi(x) = E \psi(x),
\end{eqnarray}
can be conveniently expressed in matrix form:
\begin{eqnarray}
&& \frac{d}{dx} \Psi(x) = \Lambda(x) \Psi(x), \quad \Psi(x)= \left(
\begin{array}{c}
\psi(x) \\ \psi'(x)
\end{array}
\right), \\ && \Lambda(x) = \left(
\begin{array}{cc}
0 & 1 \\ 2[V_0(x)-E] & 0
\end{array}
\right).
\end{eqnarray}
Therefore:
\begin{eqnarray}
&& \Psi(x) = b(x) \Psi(0),
\end{eqnarray}
where the $2\times 2$ {\it transfer matrix} $b(x)$ is symplectic, which implies that its
determinant is equal to $1$. The transfer matrix can be expressed in terms of two real solutions
${\rm v}_{1,2}(x)$ such that ${\rm v}_1(0)=1$, ${\rm v}_1'(0)=0$, ${\rm v}_2(0)=0$, ${\rm
v}_2'(0)=1$, as follows
\begin{eqnarray}
b(x) = \left( {\begin{array}{*{20}c}
{{\rm v}_1 (x)} & {{\rm v}_2 (x)}  \\
{{\rm v}_1 ^\prime  (x)} & {{\rm v}_2 ^\prime (x)}  \\
\end{array}} \right) .
\end{eqnarray}

The general behavior of $\psi$, and the kind of spectrum of the Hamiltonian $H_0$, depends on the
eigenvalues $\beta_\pm$ of the {\it Floquet matrix} $b(T)$, which in turn are determined by the
so-called {\it discriminant} $D=D(E) = {\rm Tr} [b(T)]$:
\begin{eqnarray}
&& \beta ^2 - D\beta + 1 = 0 \quad \Rightarrow \quad
\beta_{\pm}=\frac{D}2\pm \sqrt {\frac{D^2}4-1}, \quad \beta_+
\beta_- = 1 .
\end{eqnarray}
By picking up $\Psi(0)$ as one of the eigenvectors of the Floquet matrix $b(T)$ with eigenvalue
$\beta$, it turns out that
\begin{eqnarray}
\Psi(T) = b(T) \Psi(0) = \beta \Psi(0).
\end{eqnarray}
Moreover, for these vectors in general we have that
\begin{eqnarray}
\Psi(x+nT) = \beta^n \Psi(x). \label{blochasimp}
\end{eqnarray}
The corresponding $\psi(x)$ are known as the Bloch functions, and they exist for any $E\in{\mathbb
R}$. Depending on the value of the discriminant, the Bloch functions acquire three different
asymptotic behaviors:

\begin{itemize}

\item If $|D(E)| < 2$ for a given $E$, it turns out that
\begin{eqnarray}
\beta_\pm = \exp(\pm i k T),
\end{eqnarray}
where $k\in {\mathbb R}$ is called the {\it quasimomentum}. Equation (\ref{blochasimp}) implies
that the Bloch solutions $\psi^\pm(x)$ in this case are bounded for $\vert x\vert\rightarrow
\infty$, and hence they can be equipped with a physical meaning. Therefore, $E$ belong to an
allowed energy band.

\item The equation $|D(E)|=2$ defines the band edges, denoted
\begin{eqnarray}
E_0<E_1\le E_{1'}<E_2\le E_{2'}<\ldots <E_j\le E_{j'}<\ldots
\end{eqnarray}
For these energy values it turns out that $\beta_+ = \beta_- = \pm 1$, the two Bloch functions
tend just to one which becomes periodic or antiperiodic. Thus, these energy values belong as well
to the spectrum of $H_0$. Moreover, the Bloch eigenfunctions $\psi_{j},$ $\psi_{{j'}}$ associated
to $E_j, E_{j'}$ are real, they have the same number $j$ of nodes, and both are either periodic or
antiperiodic

\item If $|D(E)| > 2$ for a given $E$, it turns out that
\begin{eqnarray}
\beta_+ = \beta\neq 0 \in{\mathbb R}, \quad \beta_- = \frac{1}{\beta}.
\end{eqnarray}
Then, the Bloch solutions $\psi^\pm$ are unbounded either for $x\rightarrow \infty$ or for
$x\rightarrow - \infty$ so that they cannot be equipped with a physical interpretation. Therefore,
$E$ belongs to a forbidden energy gap

\end{itemize}

\begin{figure}[ht]
\centering \epsfig{file=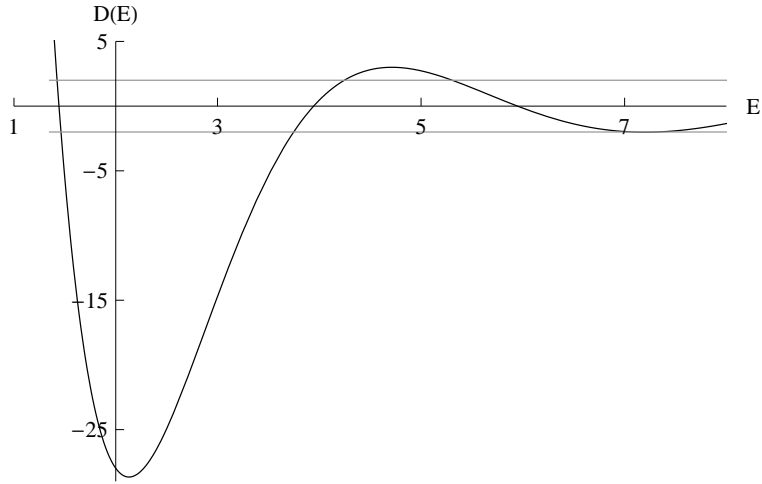, width=10cm}
\caption{The discriminant $D(E)$ versus $E$ for $V_0(x) = 5 \sin^2(x)$}
\end{figure}

As an illustration, we have plotted in Figure 6 the discriminant $D(E)$ as a function of $E$ for
$V_0(x) = 5 \sin^2(x)$.

\subsection{Lam\'e potentials}

A very special class of periodic systems are given by the Lam\'e potentials
\begin{eqnarray}
&& V_0(x) = \frac12 n(n+1) \, m \, {\rm sn}^2(x|m), \quad n\in{\mathbb N},
\end{eqnarray}
where ${\rm sn}(x|m)$ is a Jacobi elliptic function. These are doubly periodic special functions
defined by \cite{fg07}
\begin{eqnarray}
&& {\rm sn}(x|m) = \sin\varphi, \quad  {\rm cn}(x|m) = \cos\varphi,
\quad {\rm dn}(x|m) = \frac{d\varphi}{dx},
\end{eqnarray}
with $x = \int_0^\varphi \frac{d\theta}{\sqrt{1-m\sin^2\theta}}$. The real and imaginary periods
are respectively $(4K,2iK')$, $(4K,4iK')$, $(2K,4iK')$, where
\begin{eqnarray}
&& K \equiv K(m) = \int_0^{\frac{\pi}2}
\frac{d\theta}{\sqrt{1-m\sin^2\theta}}, \ K' = K(m_1), \ m_1 = 1 - m .
\end{eqnarray}

The special nature of the Lam\'e potentials is due to the following properties:

\begin{itemize}

\item They have $2n + 1$ band edges, which define $n+1$ allowed energy bands ($n$ of them finite
and $1$ infinite) and $n+1$ energy gaps ($n$ finite and $1$ infinite)

\item It is possible to determine the general solution of the Schr\"odinger equation for any real
value of the energy parameter

\item In particular, there exist explicit analytic expressions for the band edge eigenfunctions

\end{itemize}

Let us illustrate this behavior through the simplest example.

\subsection{The Lam\'e potential with $n = 1$}

In this case the Bloch functions for a real arbitrary energy value $\epsilon$ are given by
\begin{eqnarray}
&& u^{(0)^+}(x) \propto \frac{\sigma(x + \omega' + a)}{
\sigma(x+\omega')}e^{-x\zeta(a)}, \quad
u^{(0)^-}(x) \propto \frac{\sigma(x + \omega' - a)}{
\sigma(x+\omega')}e^{x\zeta(a)}, \label{blochlamen=1} \\
&& \hskip4cm  \beta = e^{2a\zeta(\omega)-2\omega \zeta(a)},
\end{eqnarray}
where $\zeta'(z)=-\wp(z), \ [\ln\sigma(z)]'=\zeta(z)$, $\omega = K, \ \omega' = i K'$ and
\begin{eqnarray}
&& 2\epsilon = \frac23 (m+1) -\wp (a) . \label{a-epsilon}
\end{eqnarray}
Note that $\wp$ is the Weierstrass $p$-function, which is related to the Jacobi elliptic function
sn through \cite{ak90}
\begin{eqnarray}
&& \wp(z) = e_3 + \frac{e_1 - e_3}{{\rm sn}^2(z\sqrt{e_1 - e_3})}.
\end{eqnarray}
There are now three band edges, and the explicit expression for the corresponding eigenvalues and
eigenfunctions are given by
\begin{eqnarray}
&& E_0 = \frac{m}2, \quad \psi^{(0)}_0(x) = {\rm dn} \, x , \\
&& E_1 = \frac12,  \quad \psi^{(0)}_1(x) = {\rm cn} \, x , \\
&& \hskip-0.5cm E_{1'} = \frac{1 + m}2,  \quad \psi^{(0)}_{1'}(x) = {\rm sn} \, x .
\end{eqnarray}
The energy axis is classified as follows: the energy interval $(-\infty,m/2)$ represents the
infinite energy gap (forbidden energies), then it comes the finite energy band $[m/2,1/2]$
(allowed energies), then it arises the finite energy gap $(1/2,1/2+m/2)$, and finally it emerges
the infinite energy band $[1/2+m/2, \infty)$.

\subsubsection{First-order SUSY using Bloch functions}

Let us choose in the first place as seed one real nodeless Bloch function $u^{(0)}$ with
$\epsilon\leq E_0$. It is straightforward to check that ${u^{(0)}}'/u^{(0)}$ is periodic, which
implies that $V_1(x)$ is also periodic (see Eq.(\ref{gv1})). The SUSY transformation maps bounded
(unbounded) eigenfunctions of $H_0$ into bounded (unbounded) ones of $H_1$. Since for $\epsilon <
E_0$ the chosen Bloch function $u^{(0)}$ diverges either for $x\rightarrow \infty$ or
$x\rightarrow -\infty$ ($\epsilon\not\in{\rm Sp}(H_0)$), then $1/u^{(0)}$ will diverge at the
opposite limits and therefore $\epsilon\not\in{\rm Sp}(H_1)$. On the other hand, for $\epsilon =
E_0\in{\rm Sp}(H_0)$, the corresponding Bloch solution $u^{(0)}$ is nodeless periodic and
therefore $1/u^{(0)}$ is also a nodeless periodic eigenfunction of $H_1$ with eigenvalue $E_0$. In
conclusion, for first-order SUSY transformations involving a nodeless Bloch function with
$\epsilon\leq E_0$ it turns out that Sp($H_1$) = Sp($H_0$), i.e., the SUSY mapping is isospectral.

In particular, for the Lam\'e potential with $n=1$ the first-order SUSY transformation which
employs the lowest band edge eigenfunction $\psi^{(0)}_0(x) = {\rm dn} \, x$ leads to
\cite{bm85,df98}
\begin{eqnarray}
V_1(x) = V_0\left(x + \frac{T}{2}\right),
\end{eqnarray}
and it is said that $V_1(x)$ is self-isospectral to $V_0(x)$. This property was discovered, for
the first time, by Braden and McFarlane \cite{bm85}, although the name is due to Dunne and
Feinberg \cite{df98}. Moreover, by using as seed the Bloch eigenfunction $u^{(0)^\pm}(x)$ with
$\epsilon < E_0$ is is straightforward to show that \cite{fmrs02a}
\begin{eqnarray}
V_1(x) = V_0(x \pm a),
\end{eqnarray}
where $a$ is related to $\epsilon$ through Eq.(\ref{a-epsilon}).

\subsubsection{First-order SUSY using general solutions}

Let us choose now a real nodeless linear combination $u^{(0)}$ of the two Bloch functions for
$\epsilon < E_0$. Although now ${u^{(0)}}'/u^{(0)}$ is not strictly periodic, it is asymptotically
periodic when $\vert x\vert\rightarrow \infty$. This property is transferred as well to the new
potential $V_1(x)$ which is asymptotically periodic for $\vert x\vert\rightarrow \infty$. As in
the previous case, the bounded (unbounded) eigenfunctions of $H_0$ are mapped into bounded
(unbounded) eigenfunctions of $H_1$. Since $\lim_{\vert x \vert \rightarrow\infty}\vert u^{(0)}
\vert = \infty$, it turns out that $1/u^{(0)}$ is square-integrable in $(-\infty,\infty)$. We
conclude that, when appropriate linear combinations of two Bloch functions of $H_0$ associated to
$\epsilon < E_0$ are used the implement the first-order SUSY transformation, we create a bound
state for $H_1$, i.e., Sp($H_1$)$= \{\epsilon\}\,\cup$\,Sp($H_0$).

We apply now this kind of transformation to the Lam\'e potential with $n=1$, using as seed a real
nodeless linear combination of the two Bloch solutions $u^{(0)^\pm}(x)$ given in
Eq.(\ref{blochlamen=1}) with $\epsilon < E_0$ \cite{fmrs02a,fmrs02b}. In this way it is obtained
an asympotically periodic potential, as illustrated in Figure 7, where the corresponding bound
state of $H_1$ is as well shown.

\begin{figure}[ht]
\centering
\epsfig{file=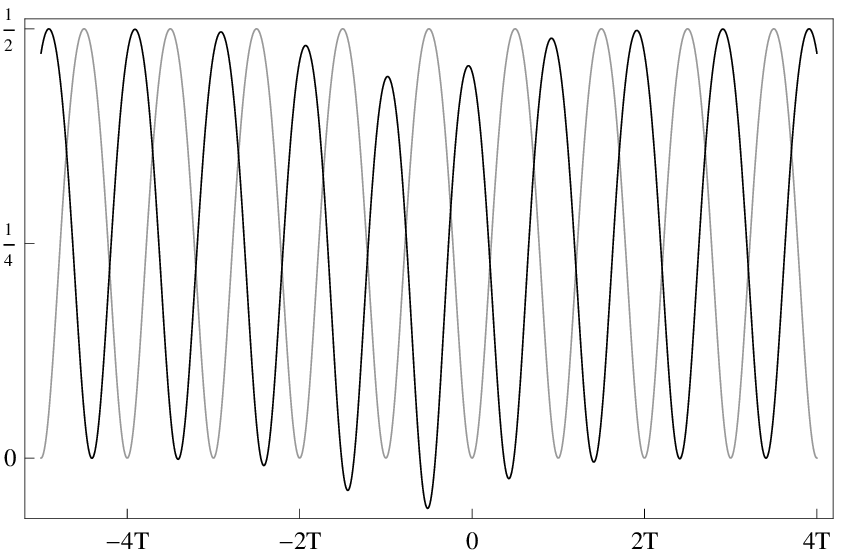, width=10cm}
\end{figure}
\begin{figure}[ht]
\centering
\epsfig{file=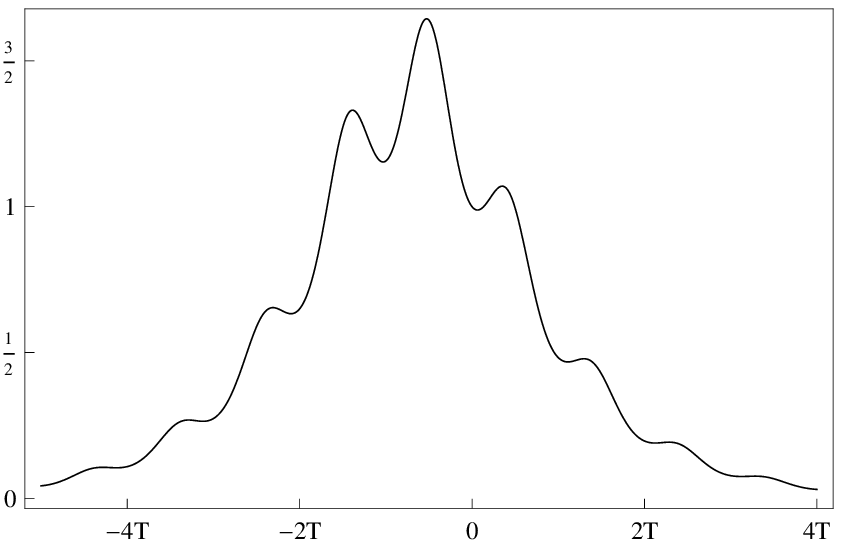, width=10cm}
\caption{Up. First-order SUSY partner $V_1$ (black curve) of the Lam\'e potential (gray curve) for
$m=0.5, \ \epsilon = 0.22$. Down: The corresponding (unnormalized) bound state of $H_1$}
\end{figure}

\subsubsection{Second-order SUSY using Bloch functions}

Let us apply now a second-order SUSY transformation using two real Bloch solutions $u^{(0)}_{1,2}$
associated to $\epsilon_{1,2}\in(-\infty,E_0]$ or $\epsilon_{1,2} \in [E_j,E_{j'}], j=1,2,\dots$
so that $W(u^{(0)}_1,u^{(0)}_2)$ is nodeless \cite{fnn00}-\cite{fmrs02b}. For this choice it turns
out that $W'(u^{(0)}_1,u^{(0)}_2)/W(u^{(0)}_1,u^{(0)}_2)$ as well as $V_2(x)$ are periodic (see
Eqs.(\ref{v2sususy},\ref{rcw2})). The SUSY transformation maps bounded (unbounded) eigenfunctions
of $H_0$ into bounded (unbounded) ones of $H_2$. Thus, $H_0$ and $H_2$ have the same band
structure, namely, Sp($H_2$)=Sp($H_0$).

For the Lam\'e potential with $n=1$, if we employ the two band edge eigenfunctions
$\psi^{(0)}_1(x),\psi^{(0)}_{1'}(x)$ bounding the finite gap, we get once again \cite{fnn00}
\begin{eqnarray}
V_2(x) = V_0\left(x + \frac{T}{2}\right).
\end{eqnarray}
Moreover, by using as seeds two Bloch functions $u^{(0)^+}_{1,2}(x)$ such that $\epsilon_{1,2} \in
(E_1,E_{1'})$ it is straightforward to show that \cite{fmrs02a}
\begin{eqnarray}
& V_2(x) = V_0(x + \delta + \delta'), \quad  \ a_1 = \delta +
\omega' , \quad a_2 = \delta' + \omega' .
\end{eqnarray}

\subsubsection{Second-order SUSY using general solutions}

Let us choose now as seeds two real linear combinations of Bloch function for
$\epsilon_{1,2}\in(-\infty,E_0]$ or $\epsilon_{1,2} \in [E_j,E_{j'}], j=1,2,\dots$ so that
$W(u^{(0)}_1,u^{(0)}_2)$ is nodeless \cite{fmrs02a,fmrs02b}. Once again, \\
$W'(u^{(0)}_1,u^{(0)}_2)/W(u^{(0)}_1,u^{(0)}_2)$ and consequently $V_2(x)$ are asymptotically
periodic for $\vert x\vert\rightarrow\infty$, the second-order SUSY transformation maps bounded
(unbounded) eigenfunctions of $H_0$ into bounded (unbounded) ones of $H_2$. Since now the two
eigenfunctions of $H_2$, $u^{(0)}_2/W(u^{(0)}_1,u^{(0)}_2)$, $u^{(0)}_1/W(u^{(0)}_1,u^{(0)}_2)$
associated to $\epsilon_1, \ \epsilon_2$, can be made square-integrable for $x\in
(-\infty,\infty)$, in turns out that Sp($H_2$)$= \{\epsilon_1,\epsilon_2\}\,\cup$\,Sp($H_0$).

Let us illustrate the procedure for the Lam\'e potential with $n=1$, using as seeds two real
linear combinations $u^{(0)}_{1,2}$ of $u^{(0)^+}_{1,2}$ and $u^{(0)^-}_{1,2}$ for
$\epsilon_{1,2}\in (E_1, E_{1'})$.  As in the general case, it is obtained a potential $V_2(x)$
which is asymptotically periodic for $\vert x\vert\rightarrow \infty$. The corresponding
Hamiltonian has two bound states at the positions $\epsilon_1,\epsilon_2$. An illustration of
these results is shown in Figure 8.

\begin{figure}[ht]
\centering \epsfig{file=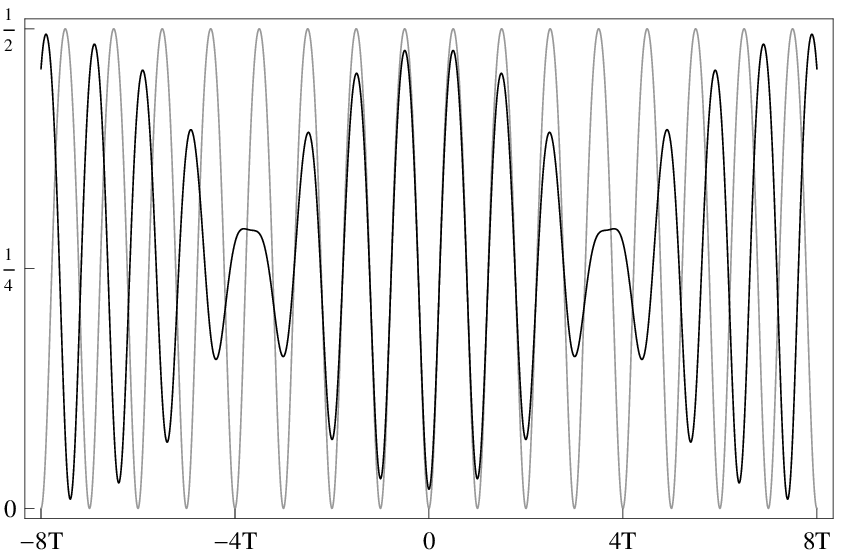, width=10cm}
\end{figure}
\begin{figure}[ht]
\centering \epsfig{file=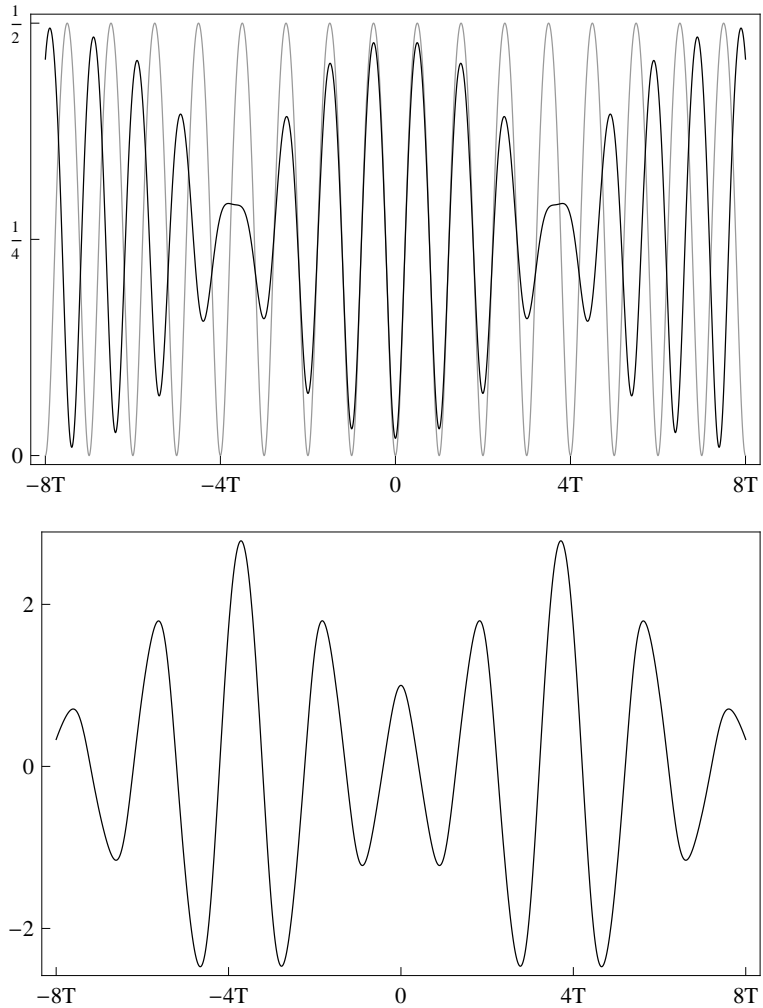, width=10cm}
\caption{Up. Second-order SUSY partner $V_2$ (black curve) of the Lam\'e potential (gray curve)
for $m=0.5, \ \epsilon_1 = 0.64, \ \epsilon_2 = 0.65$. Down: The (unnormalized) bound state of
$H_2$ associated to $\epsilon_2 = 0.65$.}
\end{figure}

Our conclusions concerning SUSY transformations applied to periodic potentials are the following.

\begin{itemize}

\item It was employed successfully the SUSY QM for generating new solvable potentials departing
from a periodic initial one

\item For seed solutions chosen as Bloch functions associated to the band edges or to
factorization energies inside the spectral gaps, the new potential is periodic, isospectral to the
initial one

\item For seed solutions which are real linear combinations of Bloch functions, the new potential
is asymptotically periodic, and the spectrum of the corresponding Hamiltonian will include a
finite number of bound states embedded into the gaps

\item The technique was clearly illustrated by means of the Lam\'e potential with $n=1$

\end{itemize}

\section{Conclusions}

\begin{itemize}

\item Along this paper it has been shown that supersymmetric quantum mechanics is a simple but
powerful tool for generating potentials with known spectra departing from a given initial solvable
one. Indeed, SUSY QM can be used to implement the spectral manipulation in quantum mechanics

\item One of our aims when writing this paper is to present the development of the factorization
method, in particular the further advances arising after Mielnik's paper about generalized
factorizations \cite{mi84}

\item Supersymmetric quantum mechanics has interrelations with several interesting areas of
mathematical physics such as solutions of non-linear differential equations \cite{fnn04,
cfnn04,mn08,dek92,vs93,ad94,srk97,acin00}, deformation of Lie algebras and coherent states
\cite{fhn94,fnr95,ro96,fh99,fhr07,ma09a,ma09b}, among others. In my opinion, a further exploration
of these interrelations is needed, which represents the future of SUSY QM. We will pursue these
issues in the next years with the hope of getting as many interesting results as we were obtaining
previously.

\end{itemize}

\bigskip

\noindent{\large\bf Acknowledgments.}
The author acknowledges Conacyt, project No. 49253-F.

\end{document}